\documentclass[vecphys]{svmult}
\usepackage{makeidx}         
\usepackage{graphicx}        
\usepackage{multicol}        
\usepackage[bottom]{footmisc}
\usepackage{natbib}          
\makeindex                   

\def\spose#1{\hbox to 0pt{#1\hss}}
\def\ltsimm{\mathrel{\spose{\lower 3pt\hbox{$\sim$}}
	\raise 2.0pt\hbox{$<$}}}
\def\gtsimm{\mathrel{\spose{\lower 3pt\hbox{$\sim$}}
	\raise 2.0pt\hbox{$>$}}}
\def\Mdot{\hbox{${\dot M}$}}

\def\km{{\rm\thinspace km}}
\def\cm{{\rm\thinspace cm}}

\def\s{{\rm\thinspace s}}
\def\yr{{\rm\thinspace yr}}
\def\kmps{\hbox{${\rm\km\s^{-1}\,}$}}

\def\pcm3{\hbox{${\rm\cm^{-3}\,}$}}

\def\Msol{\hbox{${\rm\thinspace M_{\odot}}$}}
\def\Msolpyr{\hbox{${\rm\Msol\yr^{-1}\,}$}}

\bibpunct{(}{)}{;}{a}{}{,}

\begin{document}

\title*{Mass-Loaded Flows}

\author{Julian M. Pittard}
\institute{School of Physics and Astronomy, The University of Leeds, Leeds,
LS2 9JT, UK
\texttt{jmp@ast.leeds.ac.uk}.}

\maketitle

\section{Introduction}
\begin{figure}[t]
\centering
\includegraphics[width=0.4\textwidth,angle=-90]{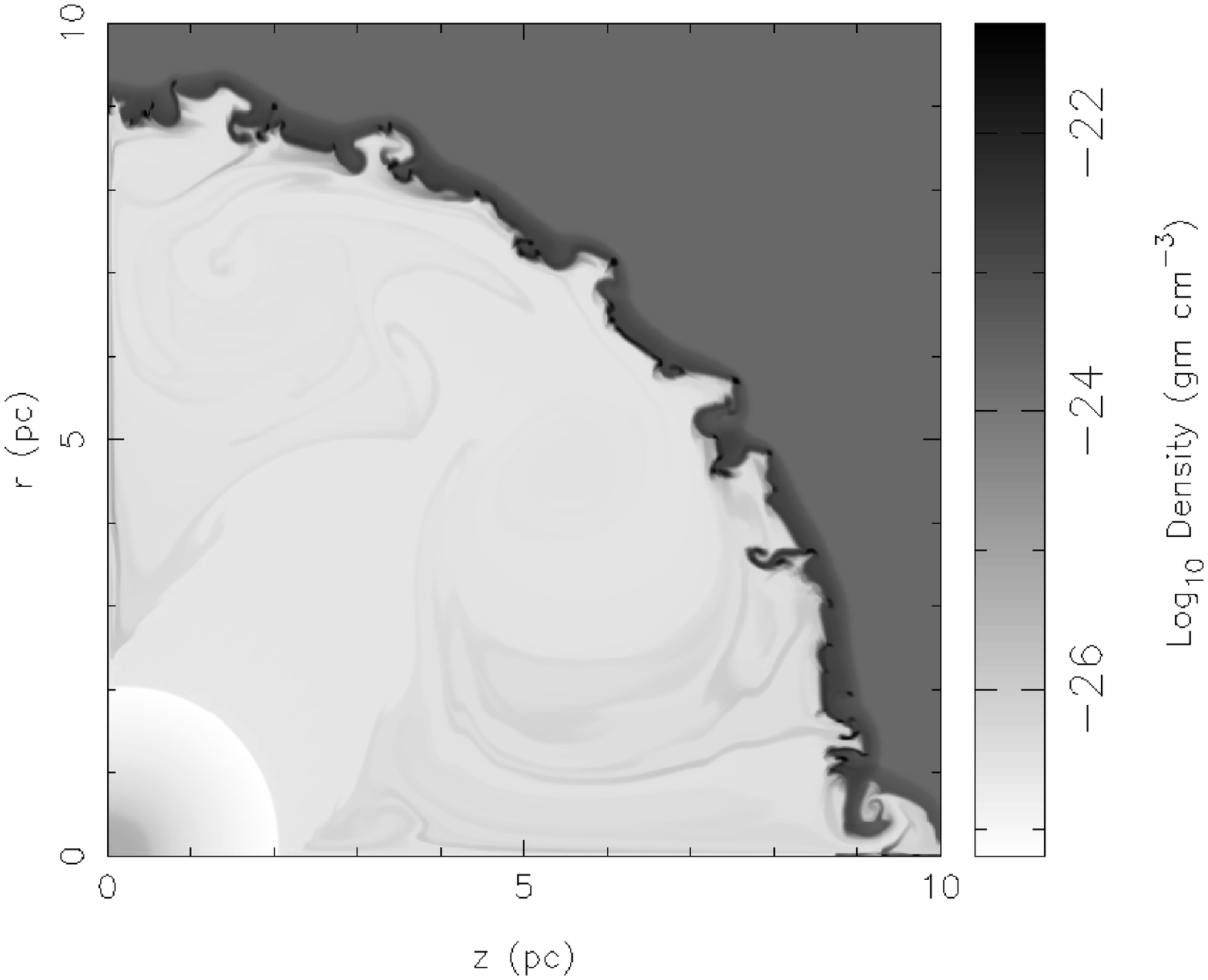}
\includegraphics[width=0.4\textwidth,angle=-90]{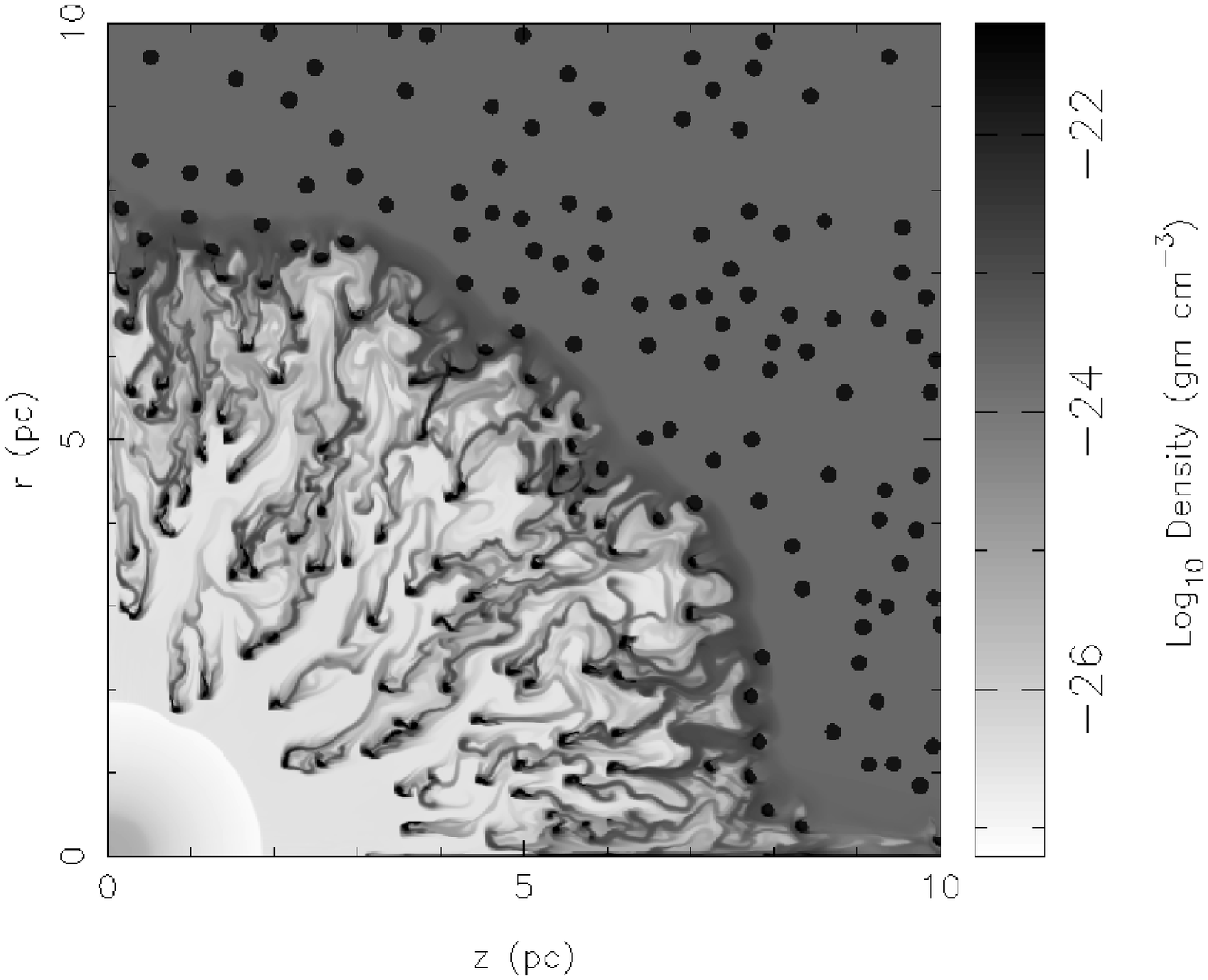}
\caption{Numerical simulations which demonstrate the effect of 
mass-loading on a `cluster-wind' from a group of
early-type stars. In the left panel the wind expands into a smooth medium,
while in the right panel the ambient medium is clumpy. The time in each 
simulation is identical.
The entrainment of mass from the clumps increases the density of the
diffuse gas in the bubble interior, reduces its temperature, and
slows its expansion.}
\label{fig:clusterwind}
\end{figure}

A key process within astronomy is the exchange of mass, momentum, and
energy between diffuse plasmas in many types of astronomical sources
(including planetary nebulae (PNe), wind-blown bubbles (WBBs),
supernova remnants (SNRs), starburst superwinds, and the intracluster
medium) and dense, embedded clouds or clumps (e.g.,
Fig.~\ref{fig:clusterwind}).. This transfer affects the large scale
flows of the diffuse plasmas as well as the evolution of the
clumps. While in much theoretical work this interaction has been
ignored, its consequences can be fundamental, as a growing body of
literature now shows. Indeed, the standard model of the interstellar
medium is based on such exchanges \citep{McKee:1977}, which occur
through, for example, conduction, ablation and photoevaporation. The
injection and mixing of mass from condensations into a surrounding
supersonic medium induces shocks, increasing the pressure of the
flowing medium \citep[e.g.,][]{Pittard:2005}.  This can lead to clump
crushing and the reduction of the Jeans mass causing star formation,
and is likely to play a role in sequential star formation
\citep[e.g.,][]{Elmegreen:1977}, and may allow a starburst to develop
\citep*{Hartquist:1997}. Radiative cooling is one way in which a
starburst might be regulated, as it acts to reduce the pressure of the
ambient medium once the mass injection rate becomes too high.

Three lengthscales characterize the entrainment of material from a
clump into a surrounding flow \citep{Hartquist:1993}. The smallest
lengthscale is associated with the turbulent boundary layer around the
clump. On intermediate scales the material injected into the flow
forms a cometary-like tail, such as those seen around clumps in
PNe. On the largest scales, the material is completely mixed into the
flow and becomes indistinguishable from it. Unfortunately, and despite
huge effort, the effectiveness of the physical processes in
controlling the interchange of dense and diffuse material remains
uncertain, in part because of the complexity of the turbulent
boundary layers which exist between them.  In addition, the
microphysics which may drive some global processes is poorly
understood. For example, magnetic reconnection, which may be
necessary in order for clump and diffuse material to fully mix, is a 
difficult subject.

The influence of John Dyson in the field of mass-loaded flows cannot
be overemphasized. He was one of the first to study the process of
photoevaporation, he was involved in the development of the widely used
analytical theory for ablatively-driven mass-loading, and in many
subsequent works he has investigated the effect of mass-loading on a
wide variety of astrophysical sources. I am grateful that I have
had the opportunity to work with him in this field.

In Sec.~\ref{sec:mass_exchange} I review our current understanding of
mass-injection processes. Sec.~\ref{sec:intermediate} focuses on
intermediate-scale structure, while Sec.~\ref{sec:global} examines the
global effect of mass-loading on a flow. Sec.~\ref{sec:mlflows}
concerns the mass-loading of a variety of diffuse sources.  For an
excellent summary of existing theoretical and observational studies on
the interface between clouds and their surroundings see
\citet{Hartquist:1993}.

\section{Mass Exchange Processes}
\label{sec:mass_exchange}
Consider a cloud of radius $r_{\rm c}$, density $\rho_{\rm c}$, and
mass $M_{\rm c} = 4/3 \pi \rho_{\rm c} r_{\rm c}^{3}$, embedded in a
medium of temperature $T$, density $\rho$, velocity $v$, and pressure
$P=\rho k T/\mu m_{\rm H}$. Let $c_{\rm c}$ and $c$ be the isothermal
sound speed in the cold cloud and in the hotter surroundings,
respectively, and $\mathcal{M}$ be the Mach number of the flow relative
to the cloud.  Mass can be lost from the cloud and entrained into the
surroundings through three main mechanisms, as discussed in the
following subsections.  I describe our current
understanding of each process, detail analytical estimates of the rate
of mass-loss, and highlight current uncertainties. The mass-loss rates
driven by each process are then compared for clumps in a variety of
different situations. 

\subsection{Hydrodynamic Ablation}
\label{sec:ab}
\begin{figure}[t]
\centering
\includegraphics[width=\textwidth]{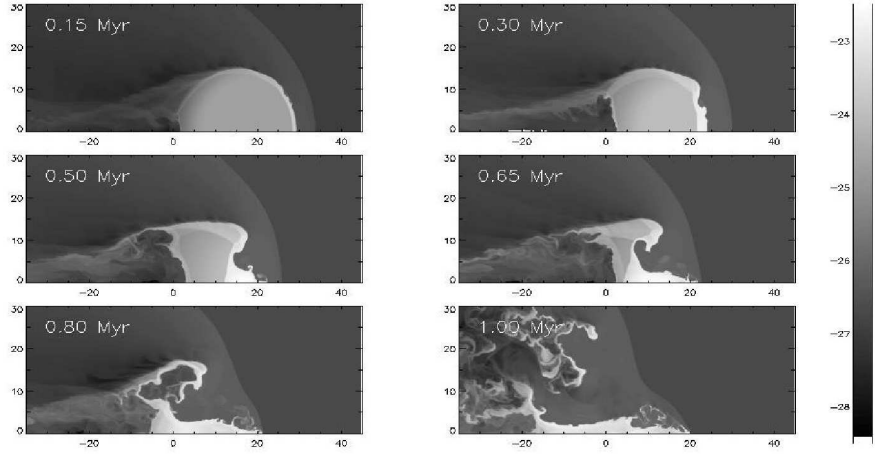}
\caption{The destruction of a cold cloud in a supersonic flow by
hydrodynamic ablation. The time evolution of the logarithm of mass
density is shown (in units of ${\rm g}\pcm3$), with distances given in
pc. At the beginning of the simulation the cloud center is at
$z=20\;$pc. $\chi=500$ and $\mathcal{M}=3$. There is no thermal
conduction or photoevaporation, magnetic fields, or self-gravity, and
the simulations are performed using 2D axisymmetry. Radiative cooling
is included. \citep[From][]{Marcolini:2005}.}
\label{fig:noheat}
\end{figure}

Numerical simulations of the interaction of a supersonic wind or a
strong shock with a single cloud have been presented many times. The
evolution for the case of an adiabatic cloud can be broken into 4
consecutive stages: an initial transient stage when the shock first
strikes the cloud, a compression stage, a re-expansion stage, and
finally a destruction stage. During the initial interaction, a bow
shock forms around the cloud, while a shock is driven into the cloud
with velocity $v_{\rm c} \approx \chi^{-1/2} v_{\rm s}$, where $\chi$
is the density ratio between the cloud and its initial (e.g.,
pre-shock) surroundings, and $v_{\rm s}$ is the velocity of the shock
through the ambient medium. The characteristic timescale for the cloud
to be crushed by the transmitted shock is $\tau_{\rm cc} =
r_{\rm c}/v_{\rm c} \approx \chi^{1/2} r_{\rm c}/v_{\rm s}$. When the
transmitted shock reaches the back of the cloud, a strong rarefraction
is reflected back into the cloud, causing its subsequent re-expansion
downstream. This is accompanied by a lateral expansion driven by the
high pressure in the cloud and the lower pressure in the surrounding
medium at its sides.  The cloud is disrupted by the action of both
Kelvin-Hemlholtz (KH) and Rayleigh-Taylor (RT) instabilities, with the
Richtmyer-Meshkov instability playing a minor role unless the surface
of the cloud is irregular. Destruction occurs after several crushing
times, with the cloud material expanding and diffusing into the
ambient flow \citep*{Klein:1994}. In 3-D simulations, instabilities
drive a richer structure \citep{Stone:1992,Xu:1995}. Recent laser
experiments confirm that the the vortex ring which forms at the back
of the cloud is broken up by the action of azimuthal bending mode
instabilities \citep{Klein:2003}.  In contrast, radiative clouds break
up into numerous dense cold fragments which survive for many
dynamical timescales \citep*{Mellema:2002,Fragile:2004}. Self-gravity
can become dynamically important in the dense fragments behind the
compression shock.  External magnetic fields generally increase the
compression of the cloud and enhance radiative cooling, while magnetic
fields internal to the cloud resist compression \citep[see][and
references therein]{Fragile:2005}.  An example of a numerical
calculation of the time evolution of a cold cloud interacting with a
supersonic wind is shown in Fig.~\ref{fig:noheat}.

An analytical theory for the hydrodynamic ablation of material from
dense clumps into the surrounding flow was presented by
\citet{Hartquist:1986}. First, consider a clump embedded in a subsonic
flow. The magnitude of the pressure variations over the surface of the
clump, created by the well-known Bernoulli effect, is
\citep{Landau:1959}
\begin{equation}
|\Delta P| \approx P_{\rm s}\left[1-\left\lbrace1+\frac{\gamma-1}{2}\mathcal{M}^{2}\right\rbrace^{-\gamma/(\gamma-1)}\right],
\end{equation}
where the stagnation pressure, $P_{\rm s} = P + \rho v^{2}$.
In the small Mach number regime, $|\Delta P|\approx
\mathcal{M}^{2}P_{\rm s}$. As the flow is fastest at the sides of the
clump, the pressure is reduced, and the clump expands in directions
normal to the flow at a speed
\begin{equation}
v_{\rm exp} \approx \frac{\gamma c_{\rm c}}{\gamma-1}\;{\rm ln}\left\lbrace1+\frac{\gamma-1}{2}\mathcal{M}^{2}\right\rbrace.
\end{equation}
For small Mach numbers, $v_{\rm exp} \approx (\gamma/2)c_{\rm
c}\mathcal{M}^{2}\approx c_{\rm c}\mathcal{M}^{2}$.  Mixing between
the cloud material and the flow occurs within a region of size $l$
which is set by the requirement that the rate of mass-loss from the
clump, $\Mdot_{\rm ab}$, is comparable to the mass-flux of the ambient
flow through this region, $\Mdot_{\rm s}$.  Since $\Mdot_{\rm ab} \sim
M_{\rm c}/t \sim M_{\rm c}v_{\rm exp}/l$, and $\Mdot_{\rm s} = \rho v
l^{2}$, $l \approx (\mathcal{M}^{2} M_{\rm c} C_{\rm c}/\rho
v)^{1/3}$.  In contrast, if the flow is supersonic, mixing occurs
largely as a result of a low pressure region over the reverse face of
the clump, `shadowed' from the flow. Since the mass cannot leave the
clump faster than its sound speed, $v_{\rm exp}\sim c_{\rm c}$, and in
this case $l \approx (M_{\rm c} c_{\rm c}/\rho v)^{1/3}$.

The rate of mass-loss from the clump, $\Mdot_{\rm ab} \approx l^{2}
\rho_{\rm l} v_{\rm exp}$, where $\rho_{\rm l}$ is the characteristic
density of ablated material at distance $l$. Momentum conservation
requires that $\rho_{\rm l} v_{\rm exp} = \rho v$, so in subsonic
flows $\Mdot_{\rm ab} \approx \mathcal{M}^{4/3} (M_{\rm c} c_{\rm
c})^{2/3} (\rho v)^{1/3}$, while in supersonic flow $\Mdot_{\rm ab} \approx
(M_{\rm c} c_{\rm c})^{2/3} (\rho v)^{1/3}$ (i.e. independent
of $\mathcal{M}$).  These estimates have received some limited support
from the numerical simulations calculated by \citet{Klein:1994}, although
the predicted scaling with the flow parameters remains to be confirmed.  

An alternative approach based on `mixing-length' theory has been
presented by \citet{Canto:1991} \citep[see also][]{Arthur:1997}. While
the boundary layer around the cloud is likely to be turbulent,
even if the cloud and the surroundings are magnetized
\citep{Hartquist:1988}, such theories are complicated by the unknown
degree to which clump gas and the tenuous plasma physically mix, and
I do not discuss them further here.

Finally, it is unclear whether the ablation process by
itself can merge the stripped material with the global flow in the
sense that its temperature, velocity, and density approach those of
the surrounding tenuous material. It may therefore be necessary to
invoke another process, such as the transfer of heat by thermal
conductivity, for the stripped material to acquire the physical state
of the surrounding medium.  Thermal conduction can accomplish this
phase transition without microscopic mixing, and acceleration to the
global flow speed is effected by the response of stripped material to
pressure gradients and viscous coupling, which may arise from a host
of mechanisms including turbulence.

\subsection{Conductively-Driven Thermal Evaporation}
\label{sec:cond}
Cold clouds embedded in a hot medium may also lose mass to their
surroundings as hot electrons deposit energy in the surface regions of
the clump. This process is referred to as thermal evaporation. The
rate of mass loss is dependent on many factors, including the
temperature of the hot phase, the clump radius, whether the
conductivity is saturated, the presence of magnetic fields and plasma
instabilities, and whether there is a velocity difference between the
ambient medium and the cloud \citep[e.g.,][]{Cowie:1977,McKee:1977b}.
Nonspherical clumps may be treated in an approximate way by adopting
half the largest dimension as the radius of the clump
\citep{Cowie:1977b}.

If the mean-free-path for electron-electron collisions, $\lambda_{\rm
ee}$, is approximately less than the temperature scale-height,
$T/|\nabla T|$, then the heat flux into the cloud, $q$, is given by
the classical theory of thermal conduction i.e. $q = q_{\rm cl} =
-\kappa \nabla T$.  The mean-free-path $\lambda_{\rm ee}=t_{\rm
ee}(3kT/m_{\rm e})^{1/2}$, where the electron-electron
equipartition time is given by \citep{Spitzer:1962}
\begin{equation}
t_{\rm ee} = \frac{3 m_{\rm e}^{1/2} (kT)^{3/2}}{4 \pi^{1/2}
n_{\rm e} e^{4} \;{\rm ln \;\Lambda}},
\end{equation}
where ${\rm ln\;\Lambda}=29.7 + {\rm ln}\;(T/10^{6}\sqrt{n_{\rm e}})$
is the Coulomb logarithm and the other symbols have their usual
meaning. I have implicitly assumed that $T_{\rm e}=T$. 
The thermal conductivity, $\kappa$, in a fully ionized
hydrogen plasma is \citep[see, e.g.,][]{Spitzer:1962,Cowie:1977}
\begin{equation}
\kappa = 1.84 \times 10^{-5}\frac{T^{5/2}}{{\rm ln}\;\Lambda} 
\;\;{\rm erg\;s^{-1}\;K^{-1}\;cm^{-1}},
\end{equation}
\citep[the zero current requirement reduces the effective coefficient
of conductivity by a factor $\approx 0.4$ - see][]{Spitzer:1953}.  The
evaporative mass-loss rate from a single clump \citep{Cowie:1977} is
then
\begin{equation}
\label{eq:mdot_evap}
\Mdot_{\rm con} = \frac{16 \pi \mu \kappa \omega r_{\rm c}}{25 k} = 2.75\times10^{19}\;\omega \; r_{\rm pc} \; T_{6}^{5/2} \;\;{\rm g\;s^{-1}} 
\end{equation}
where $r_{\rm pc}$ is the clump radius in parsecs, and 
$T_{6} = T/10^{6}\;{\rm K}$. For classical evaporation, $\omega=1$.
As conductively driven evaporation has a very temperature sensitive
rate, ablation is likely to regulate clump dispersal in lower
temperature media.

When $\lambda_{\rm ee} \gtsimm T/|\nabla T|$, the classical theory of
thermal conduction, which is based on a diffusion approximation, may
no longer be used. Instead the heat flux reaches a limiting value;
i.e. it becomes saturated.
The approach to saturated conduction is still partly empirical, and it
is common practice to take a flux-limited form for the heat flux:
$q_{\rm sat} = 5 \phi \rho c^{3}$ \citep{Cowie:1977}, where $\rho$ and
$c$ are in the hot phase, and $\phi$ is a parameter of order unity
which describes the uncertainty in the numerical value of the
saturated heat flux. Observations of the highly saturated solar wind,
laboratory plasma experiments, and Fokker-Planck calculations suggest
that $0.3 \ltsimm \phi \ltsimm 1.1$ \citep{Giuliani:1984}.
\citet{Balbus:1982} conjectured that
the effective heat flux, $q$, can be approximated by the harmonic mean of 
$q_{\rm sat}$ and $q_{\rm cl}$,
\begin{equation}
\frac{1}{q} \approx \frac{1}{q_{\rm sat}} + \frac{1}{q_{\rm cl}}.
\end{equation}
The resulting heat flux reduces to the smaller of the two conduction forms
when there is a large disparity between them, and has the convenient 
property of a smooth transition from diffusive to flux-limited transport.
The ratio of classical to saturated heat flux is
\begin{equation}
\sigma = \frac{q_{\rm cl}}{q_{\rm sat}} = \frac{\kappa}{5 \phi \rho c^{3}}\frac{dT}{dr},
\end{equation}
where the last expression explicitly assumes spherical symmetry
i.e. $\nabla T = dT/dr$. The heat flux is then
\begin{equation}
\label{eq:q_eff}
q = \frac{\kappa}{1 + \sigma}\frac{dT}{dr},
\end{equation}
where $\sigma$ is the {\em local} stauration parameter.  This
expression becomes the diffusive flux when $\sigma \ll 1$, and the
saturated flux when $\sigma \gg 1$.
 
\citet{Cowie:1977} also defined a {\em global} saturation parameter,
\begin{equation}
\label{eq:sigma_0}
\sigma_{0} \equiv \frac{2 \kappa T}{25 \phi \rho c^{3} r_{\rm c}} = \left(\frac{T}{1.54 \times 10^{7}}\right)^{2} \frac{1}{n r_{\rm pc} \phi},
\end{equation}
where ${\rm ln\;\Lambda}=30$ has been assumed.
Whereas $\sigma$ measures the saturation locally, $\sigma_{0}$
measures global scales and allows a quick assessment of the importance
of saturation effects ($\sigma_{0}$ is essentially the local
saturation parameter, $\sigma$, evaluated at the ambient conditions
with $dT/dr = T/r_{\rm c}$).  For $\sigma_{0} \ltsimm 0.03/\phi$,
radiative losses quench the evaporation, and the clump grows in mass
as surrounding material condenses onto it \citep{McKee:1977b}. This,
of course, may make the clump gravitationally unstable, and initiate
new star formation. For $0.03/\phi \ltsimm \sigma_{0} \ltsimm 1$, the
clump is evaporated at the classical rate. The
onset of saturation occurs when $\sigma_{0}$ is of order unity, with
highly saturated flows having $\sigma_{0} \gg 1$. The
mass-loss rate in the saturated regime is specified with $\omega
\approx (1 + \sigma_{0})^{-0.7}$ \citep{Giuliani:1984}.
Since the onset of saturation is dependent on the radius of the clump
for a specified hot phase, larger clumps will tend to evaporate in the
classical limit, while the evaporation of mass from smaller clumps
will tend towards saturation. If the temperature and density of the
hot phase is evolving (e.g., because it is the interior of a WBB or
SNR), the radius of clumps which are just at the onset of saturation
will also change. 

The dynamics of the evaporation process have been analyzed by
\citet{McKee:1975} and \citet{Cowie:1977}, and an analogy with
ionization fronts can be made. For classical evaporation (i.e. $0.03
\ltsimm \sigma_{0}/\phi \ltsimm 1$), the velocity of the conduction
front, $v_{\rm cond} \sim 2 \sigma_{0} \phi c_{\rm c}^{2}/c$.  If
$\sigma_{0} \gtsimm 0.25/\phi$, the conduction front drives a shock
into the cloud. Otherwise, the cloud evaporates
subsonically. Conduction fronts in the saturated regime have a
velocity $v_{\rm cond} \sim 1.12 \sigma_{0}^{1/8}c_{\rm c}^{2}/c_{\rm
w}$ (for $\phi=1$), and always drive a shock into the cold cloud
ahead of the conduction front itself.
 
Time-dependent hydrodynamical simulations of clouds overrun by a
strong shock and undergoing conductively-driven thermal evaporation in
the classical regime have been calculated by
\citet{Orlando:2005}. Conduction inhibits the growth of
Rayleigh-Taylor and Kelvin-Helmholtz instabilities, and the
fragmentation of the cloud, but heats the evaporated material so that
it quickly becomes part of the ambient flow. Simulations in the
saturated regime confirm that the conduction front initially drives a
shock into the cloud \citep{Ferrara:1993,Marcolini:2005},
substantially increasing its density. The resulting evolution then
appears sensitive to assumptions concerning the cooling, with the
cloud and the evaporation either settling into a quasi-steady-state
\citep{Ferrara:1993}, or displaying oscillatory behaviour
\citep{Marcolini:2005}. The mass-loss rate during the initial phase is
comparable to that inferred from Eq.~\ref{eq:mdot_evap}, but it is
substantially reduced as the cloud is compressed and decreases in size
\citep{Marcolini:2005}.  Somewhat different models including
self-gravity have been calculated by \citet{Vieser:2000} and
\citet{Hensler:2002}.

\begin{figure}[t]
\centering
\includegraphics[width=\textwidth]{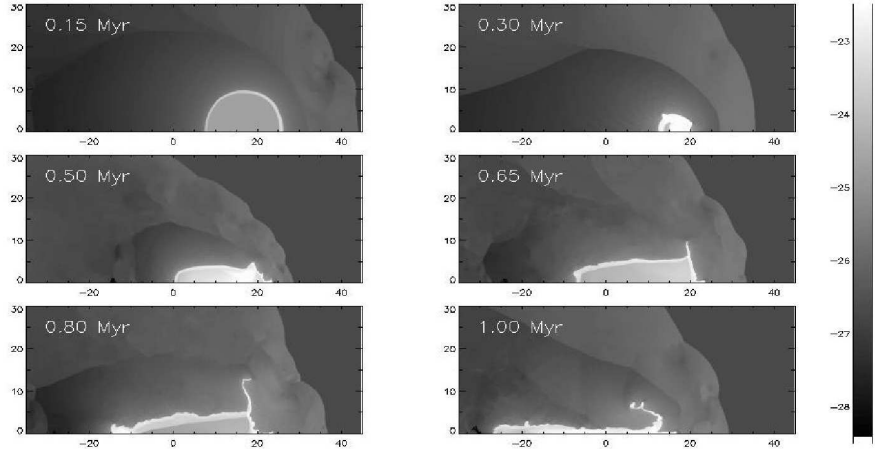}
\caption{As Fig.~\ref{fig:noheat}, but with heat conduction in the saturated
regime ($\sigma_{0} = 16$). A strong shock is driven into the cloud ahead of
the conduction front and compresses it from all sides, so that its radius
has decreased by a factor of 3 by $t=0.3\;$Myr. The subsequent evolution is
also markedly different from the purely hydrodynamical case.
 \citep[From][]{Marcolini:2005}.}
\label{fig:heat}
\end{figure}

Conductively-driven evaporation is perhaps the best-studied of the three
processes highlighted in this section, but
there are uncertainties in many physical processes whose influence on
$\Mdot_{\rm con}$ remains poorly quantified. For instance, when
$\sigma_{0} \gtsimm 100$, viscous stresses have the potential to
increase $\Mdot_{\rm con}$ significantly, but the exact enhancement
depends on the uncertain degree to which they might also saturate
\citep{Draine:1984}. On the other hand, if the electron mean free path
is reduced, the heat flux will be inhibited.
\citet{Bandiera:1994a,Bandiera:1994b} have emphasized that when
$\sigma_{0}\gtsimm 100$, the requirement of a zero net current means
that hot electrons are stopped by electrostatic effects within a thin
surface layer. The resulting heat flux is then considerably lower than
that obtained if the electrons penetrate deep into the cloud and
directly heat it through Coulomb collisions \citep{Balbus:1982}.
The electron mean free path can also be reduced
as a result of scattering by plasma instabilities, such as the
ion-acoustic instability \citep{Galeev:1984}, and by whistler waves
\citep{Gary:1977,Levinson:1992}. However, it is difficult to obtain a
self-consistent model of these processes, and the presence of a strong
magnetic field may suppress such instabilities. Partial ionization and
non-equilibrium cooling are other possibilities for suppressing the
conductivity \citep{Bohringer:1987}. Magnetic fields may not reduce the 
conductivity as much as previously thought 
\citep[e.g.,][]{Balbus:1986,Rosner:1989,Cho:2003} unless the cloud is
magnetically disconnected from its surroundings. An example of a numerical
calculation of the time evolution of a conductively-evaporating cold cloud 
interacting with a hot supersonic wind is shown in Fig.~\ref{fig:heat}.

\subsection{Photoevaporation}
\label{sec:photoevap}
The fate of a neutral clump exposed to a strong ionizing radiation
field has been extensively studied over the years. Small, low-mass
clumps are instantly ionized and rapidly dissipate i.e. are
`zapped'. In contrast, clouds which are sufficiently dense and large 
trap the ionization front, which becomes D-critical
\citep{Dyson:1968,Bertoldi:1989} and moves into the cloud driving a
shock front ahead of it. The ionized gas streams away perpendicular to
the ionization front and expands supersonically into the interclump
medium, reaching an asymptotic Mach number of $\approx 2$
\citep[e.g.,][]{Kahn:1969}. This flow may absorb a large part of the
incident ionizing flux and appear as a bright rim to the clump. 
The shock driven into the clump is focussed onto the clump axis, and can
substantially increase the initial clump density
\citep*{Sandford:1982}. This `radiation-driven implosion' may make the
clump gravitationally unstable and lead to new star
formation. Otherwise, the pressure overshoot causes the clump to
reexpand, and it undergoes several radial oscillations
\citep{Lefloch:1994} before obtaining an equilibrium, cometary-shaped
structure \citep{Bertoldi:1990}. Recombination may occur in the shadow
of the clump, but this is prevented if there is a diffuse component to
the radiation field \citep{Pavlakis:2001}. The pressure in the
evaporating flow declines rapidly, and eventually a termination shock
forms. If the surrounding medium is supersonic a bow-shock is also
formed. Photoevaporated flows also occur from the neutral disks which
surround pre-main-sequence stars, known as `proplyds' \citep*{ODell:1993,
Bally:1998}.

Analytical equations for the mass injection rate of the
photoevaporated flow as the clump is destroyed are presented in
\citet{Bertoldi:1989} and \citet{Mellema:1998}. Good agreement with
results from numerical simulations is obtained. However, it is
possible to obtain a simple estimate by setting $\Mdot_{\rm ph} =
mFA$, where $m$ is the mass per particle of the neutral material, and
$F$ and $A$ are respectively the rate per unit time per unit area at
which hydrogen ionizing photons reach the ionization front and its
area.  $F$ is reduced by absorptions in the photoevaporating flow, and
is approximately given by \citep{Mellema:1998}
\begin{equation}
F \approx \frac{F_{0}}{\left(1 + \alpha_{\rm B}F_{0}r_{\rm if}/3c_{\rm i}^{2}\right)^{1/2}},
\end{equation}
where $\alpha_{\rm B}=2.6\times10^{-13}\cm^{3}\s^{-1}$ is the case B
recombination rate for H, $r_{\rm if}$ is the radius of curvature of
the ionization front, $c_{\rm i}$ is the isothermal sound speed of the
ionized gas, and $F_{0} = \dot{S}/4 \pi d^{2}$ is the flux delivered
by the ionizing sources at the position of the clump. Typical
assumptions are $r_{\rm if} \approx r$ and $A \approx \pi r^{2}$.
$\Mdot_{\rm ph}$ declines with time as the clump is destroyed and $r$
decreases.

Photoevaporation may be suppressed if the ram or thermal pressure of
the surrounding medium is greater than the pressure of the evaporating
flow \citep{Dyson:1994}. Density inhomogeneities within the clump may
affect the process of photoevaporation, as may the instability of
inclined ionization fronts \citep{Williams:2002}. The role of magnetic
fields on the structure of ionization fronts \citep{Williams:2000} may 
also affect $\Mdot_{\rm ph}$.


\subsection{Comparison of Mass-Injection Rates}
\label{sec:mdot_comp}
While the exact rates of mass-loss by the three processes described in
Sec.~\ref{sec:ab}-\ref{sec:photoevap} remain somewhat uncertain, an 
appreciation of their relative importance can be obtained by considering 
several different objects.

\subsubsection{The Helix Nebula (NGC~7293)}
The Helix Nebula is famous for containing many extended cometary
tail-like structures, which emanate from dense, neutral globules, and
point away from the central star. The clumps have on average the
following properties: $r_{\rm c} \approx 10^{-3}\;{\rm pc}$, $T_{\rm
c} \approx 10\;{\rm K}$, $n_{\rm c} \approx 10^{6} \;\pcm3$
\citep{Dyson:1989}. A typical distance of a clump from the central
ionizing star is $d \approx 0.1\;{\rm pc}$. The star has an ionizing
photon flux $\dot{S}_{49} = \dot{S}/10^{49} \approx 3.5 \times
10^{-4}$.  The clumps appear to be overrun by [OIII] gas with $T
\approx 10^{4}\;{\rm K}$, $n \approx 10^{3}\;\pcm3$, and a relative
velocity, $v \approx 17\;\kmps$ \citep{Meaburn:2005}. Hence, the Mach
number of the flow relative to the clumps is about 1.5. Using the
analytical equations in Secs.~\ref{sec:ab} through to~\ref{sec:photoevap}, we
determine that $\Mdot_{\rm ab} \approx 1.6 \times 10^{16} \;{\rm
g\;s^{-1}}$ and $\Mdot_{\rm ph} \approx 2.4 \times 10^{17} \;{\rm
g\;s^{-1}}$. The global saturation parameter, $\sigma_{0} \approx 4
\times 10^{-7}$ for $\phi=1$, and gas would like to condense onto the
clumps at a rate $\Mdot \approx 2 \times 10^{16} \;{\rm g\;s^{-1}}$
\citep{Cowie:1977}, though this is likely prevented by the mass-loss
that occurs through photoevaporation and ablation. The estimated
lifetime of such clumps is in excess of $4 \times 10^{4}\;{\rm yr}$,
compatible with the estimated age of the nebula. The origin of the
knots is discussed in \citet{Dyson:1989}.

\subsubsection{The Wolf-Rayet Nebula RCW\thinspace58}
RCW\thinspace58 is a nebula surrounding the Wolf-Rayet star WR\thinspace40, 
and which has been formed by the current stellar wind sweeping up wind material
from previous evolutionary stages, some of which is clumpy
\citep{Chu:1982,Smith:1988}. The clumps are ionized by the central
star and typically have the following properties: $r_{\rm c} \approx
0.1\;{\rm pc}$, $T_{\rm c} \approx 10^{4}\;{\rm K}$, $n_{\rm c}
\approx 10^{3} \;\pcm3$ \citep*{Arthur:1996}. They are embedded in
shocked stellar wind material, with $n \approx 1\;\pcm3$, $v \approx
200 \;\kmps$. The flow around the clumps has $\mathcal{M} \approx
0.6$. The H-ionizing flux from the star is $\dot{S}_{49}\approx 2.5$
(P.~Crowther, private communication).  For clumps at a distance of
1~pc from the star, $\sigma_{0} \approx 4$, and the
conductively-driven mass-loss is mildly saturated with $\Mdot_{\rm con}
\approx 3 \times 10^{20}\;{\rm g\;s^{-1}}$.  A higher mass-loss rate
is obtained for ablation: $\Mdot_{\rm ab} \approx 10^{21}\;{\rm
g\;s^{-1}}$. Since the cloud is already ionized it does not make sense
to calculate a photoevaporation rate, but this may once have been the
dominant process if the cloud was formerly neutral.  Therefore,
ablation would appear to control the rate at which mass is currently stripped
from the clump. 

\subsubsection{Within a WBB}
\begin{figure}
\centering
\includegraphics[width=\textwidth]{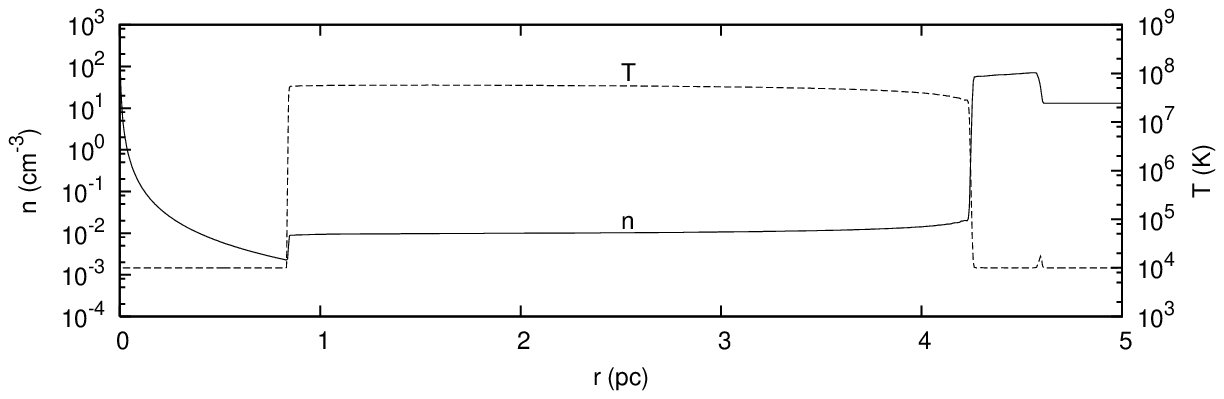}
\includegraphics[width=\textwidth]{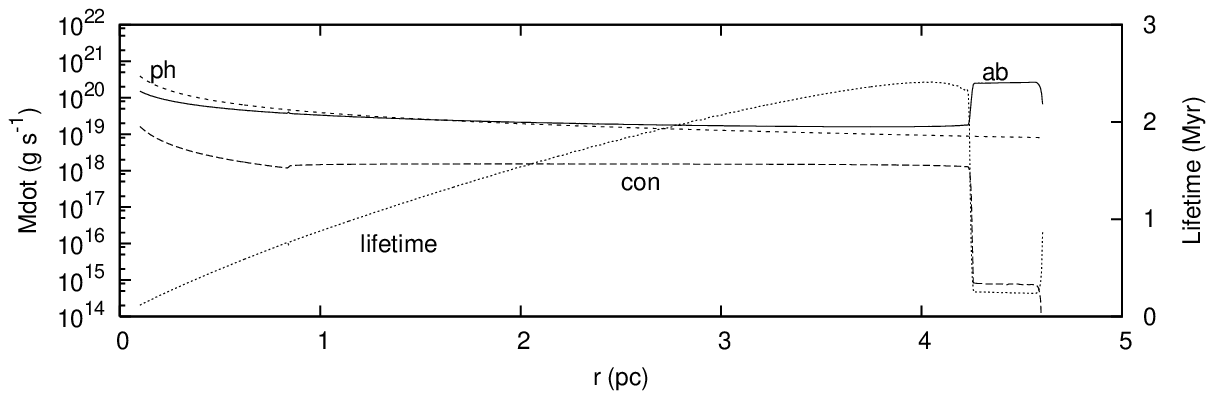}
\caption{Top: The internal structure (total number density and
temperature) of a WBB of age 0.1~Myr expanding into a stationary
medium with $n_{\rm H}=10\;\pcm3$. The central star has $\Mdot =
10^{-6}\;\Msolpyr$, $v_{\infty} = 2000\;\kmps$, and $\dot{S}_{49} =
1.2$. Bottom: The mass-loss rate from clumps of radius $0.01\;{\rm
pc}$, $n = 10^{7}\;\pcm3$, $T=100\;{\rm K}$, due to hydrodynamic
ablation (ab), photoevaporation (ph), and conductively-driven
evaporation (con). No clumps are assumed to reside within the central
0.1~pc of the bubble. Clumps located within the hypersonic stellar
wind are assumed to be surrounded by a sheath of hot gas bounded by a
bowshock on their windward surface. The conductively-driven
evaporation is highly saturated for clumps within the hot gas of the
bubble given the paramaters chosen, but is in the condensational
regime within the shell of swept-up ambient material. Also shown is an
estimate for the lifetime of the clumps: $t = (\Mdot_{\rm ab}+\Mdot_{\rm
con}+\Mdot_{\rm ph})/M_{\rm c}$).}
\label{fig:mdot_bub}
\end{figure}

Of course, the mass-loss rate from each process varies within a bubble
as the density, velocity, temperature, and distance from the ionizing source
change. This is demonstrated in Fig.~\ref{fig:mdot_bub}, where the top panel
shows an example of the density and temperature structure within a bubble,
while the mass-loss rates and lifetime of a specific clump are shown in the
bottom panel. 

\section{Intermediate-Scale Structure}
\label{sec:intermediate}
\begin{figure}
\centering
\includegraphics[width=3.2cm]{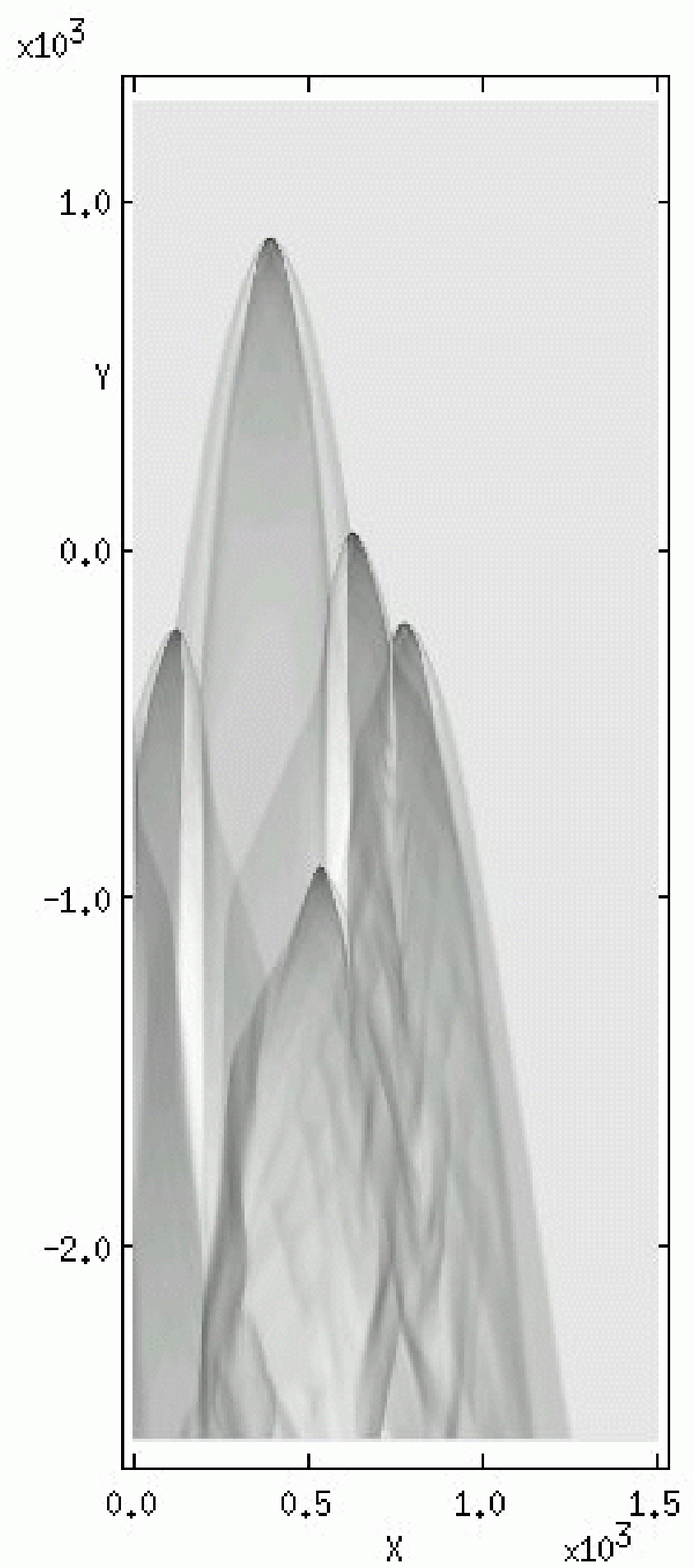}
\includegraphics[width=6.35cm]{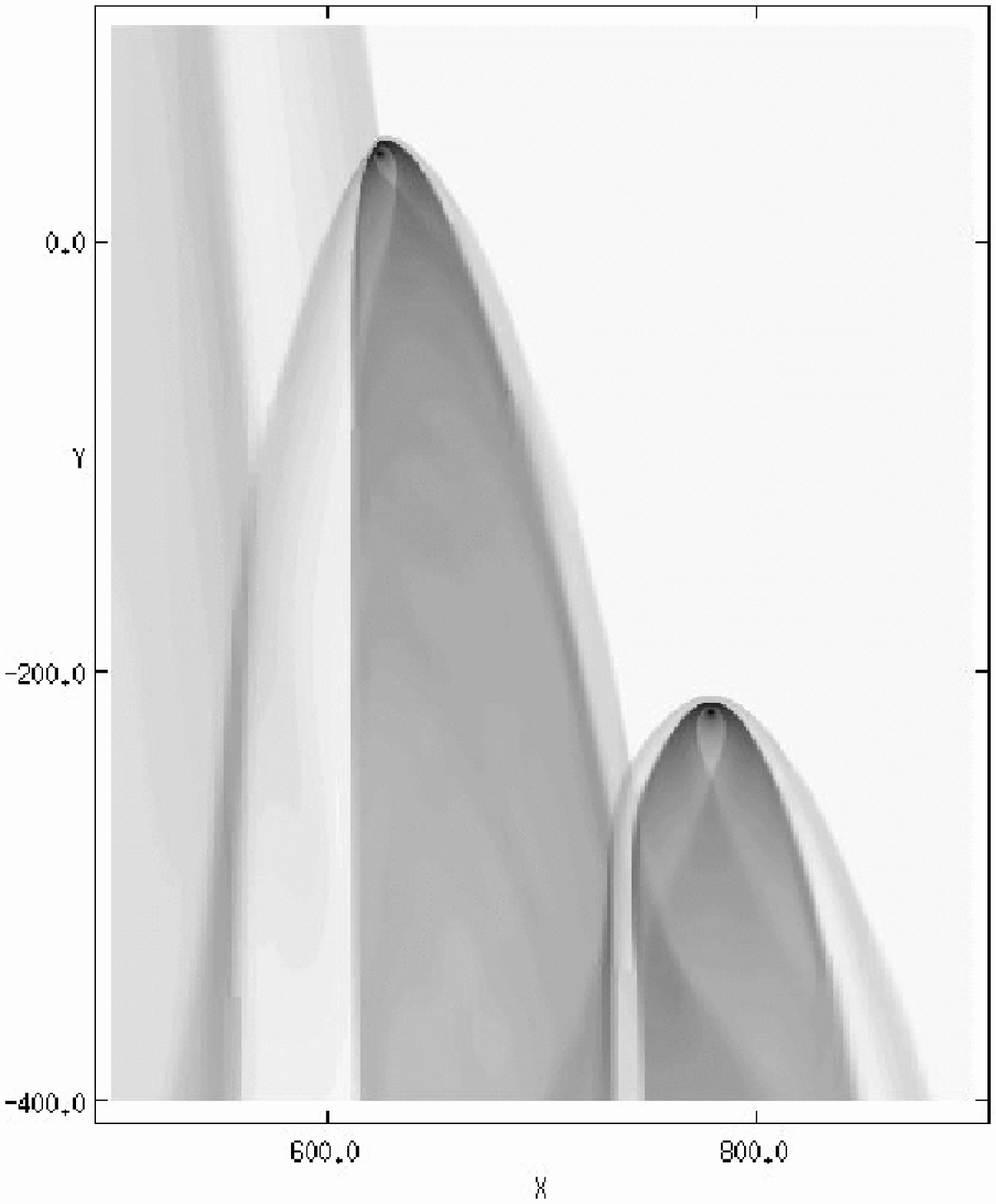}
\caption{The interaction of a hypersonic flow with multiple cylindrical mass
sources when the combined mass injection rate is comparable to the mass
flux of the incident flow through the volume that the mass sources occupy
\citep[from][]{Pittard:2005}. The right panel shows an enlargement of the
flow around two of the mass sources.}
\label{fig:intermediate}
\end{figure}

While the stripping of mass from clumps has been extensively, though
not definitively studied, to date there has been very little
work on how intermediate-scale structures disperse/merge into the
background flow. This problem has received some attention in studies
of the interaction between multiple clouds and a flow
\citep*[e.g.,][]{Jun:1996,Poludnenko:2002,Steffen:2004}, but our
understanding of such interactions is still developing.  A limitation
of these works is that the clumps have been modelled as single-phase
entities, and the density contrast between the clumps and their
surroundings has, for numerical reasons, typically been $\sim
10^{2}$. The simulated clumps then have such short lifetimes that they
are unable to significantly mass-load the flow. In reality, in many
astropysical flows the density contrasts are much larger. A different
numerical approach which accounts for the much longer lifetimes of
clumps is to set up sources which continuously inject mass into the
flow \citep{Falle:2002}.  Multiple sources act as an efficient barrier
to the flow if they are sufficiently close together that their
combined mass injection rate is comparable to or exceeds the mass flux
of the incident flow into the volume that they occupy
\citep[][see also Fig.~\ref{fig:intermediate}]{Pittard:2005}. 
In such cases, the thermal pressure of the flow
is greatly enhanced (at the expense of its ram pressure), and
crucially becomes relatively uniform - these conditions are exactly those
required to increase the probability of cloud collapse and new star
formation. 

Perhaps one of the best examples of intermediate scale structure is
the comet-like tail extending from the Galactic Centre source IRS~7, a
red supergiant \citep*{Yusef-Zadeh:1991,Serabyn:1991}. Although
originally interpreted in terms of a global wind-wind collision
between the slow dense wind from IRS~7 and a Galactic wind,
this model had difficulty in explaining the length of the tail
\citep{Yusef-Zadeh:1992}. However, it may be overcome if the wind of
IRS~7 is clumpy enough that it becomes semi-porous to the Galactic
wind, with much smaller bow-shocks forming around each clump
\citep{Dyson:1994b}.

\section{The Global Effect of Mass-Loading on a Flow}
\label{sec:global}
The effect of mass-loading on a global flow can be studied in a
one-fluid approximation if the following two assumptions are made: i)
the clumps are sufficiently numerous that they can be considered
to be continuously distributed; ii) the characteristic scale
length of injection and mixing is much smaller than the dimensions of
the global flow (so that the injected mass reaches the general flow
velocity and temperature essentially instantaneously).  For the steady
injection of mass from an ensemble of clumps into the interior of a
spherically expanding flow, the time-independent continuity and momentum 
equations are then
\begin{equation}
\frac{1}{r^{2}}\frac{\partial(r^{2} \rho u^{2})}{\partial r} = Q
\end{equation}
and
\begin{equation}
\rho v \frac{\partial v}{\partial r} + \frac{a^{2}}{\gamma}\frac{\partial \rho}{\partial r} + \frac{2 \rho a}{\gamma}\frac{\partial a}{\partial r} = - Qv,
\end{equation}
where $r$ is the distance from the star, and $a$ is the isothermal
sound speed of the flow. $Q$ is the mass loaded into the flow per unit
time per unit volume. The clumps are assumed, on average, to be
stationary with respect to the global flow, so that there is no net rate
of momentum injection.

The continuity and momentum equation may be combined to give
\begin{equation}
(v^{2} - a^{2}) \frac{\partial v}{\partial r} = - \frac{Q}{\rho}(v^{2}+a^{2})
+ \frac{2 a^{2} v}{r}.
\end{equation}
If $r$ is large (i.e. the flow is plane-parallel) and if $v > a$
(i.e. $M > 1$), then $v^{2} > a^{2}$, $\partial v/\partial r < 0$, and
the flow decelerates. On the other hand, if $r$ is large and $v < a$
(i.e. $M < 1$), then $v^{2} < a^{2}$, $\partial v/\partial r > 0$, and
the flow accelerates. A key feature of mass-loading is that it tends
to drive the ambient flow to Mach number unity \citep{Hartquist:1986}.
For an expanding, time-dependent flow, mass-loading may drive the Mach
number to 0.6-0.7 \citep*{Arthur:1993}.  Another feature is that if
mass-loading is large, a global reverse shock will be greatly weakened 
\citep*{Arthur:1994,Williams:1995,Williams:1999,Pittard:2001b}. In addition, 
a two-fluid study has revealed that shocks which overrun clumpy media
are weakened and broadened \citep{Williams:2002b}. This has
significant implications for the survival of the clumps.
Finally, the injection of mass into a radiative medium has a stabilizing
influence against isobaric and isentropic perturbations, and can suppress
the development of the thermal instability \citep{Pittard:2003}.

\section{Sources with Large-Scale Mass-Loaded Flows}
\label{sec:mlflows}
Theoretical studies of the effect of mass-loading on a flow fall into
two categories. If the flow is treated as a single fluid and has a
very simple geometry, it is often possible to obtain a similarity
solution.  Alternatively, one can use a hydrodynamical code to model
and find a wider variety of solutions.  This allows much greater
flexibility, and the ability to incorporate a larger number of
physical processes.  In principle, a two-fluid approach can be used,
in which a time-dependent spectrum of clumps is separately tracked. Of
course, an unavoidable drawback of such computations is the loss of
generality, as, for instance, the clump spectrum inside a starburst
superwind is likely to be somewhat different to that in our local ISM.

In the following subsections I discuss mass-loading in a variety of
astrophysical settings. The properties of winds mass-loaded by
material from clumps (or stellar sources) has also received
theoretical attention in the context of ultracompact H~II regions
\citep*{Dyson:1995}, Herbig-Haro objects \citep[see discussion
in][]{Hartquist:1988,Hartquist:1993}, and globular cluster winds
\citep{Durisen:1981}. The mass-loading of accretion flows in stellar 
clusters and AGNs has also been studied by \citet{David:1989} and
\citet*{Toniazzo:2001}.

\subsection{WBBs, PNe, and Superbubbles}
The dramatic and beautiful impact of stellar winds on their
environment has long been recognized (e.g., see the chapter by Jane 
Arthur). 
The first similarity solutions of this interaction,
which assumed a smooth wind and environment, were obtained by
\citet{Dyson:1972} and \citet{Dyson:1973}. The shocked ambient gas was
found to cool very rapidly \citep[as subsequently shown in a numerical
simulation by][]{Falle:1975}, and the hot shocked wind was found to fill
most of the bubble volume. The effect of mass-loading was first
considered by \citet{Weaver:1977}, who obtained an approximate
similarity solution for a WBB expanding into a constant density medium
by assuming an isobaric shocked wind region. In this work,
mass-loading of the interior of the WBB is assumed to occur through
the conductively-driven evaporation of the cool swept-up shell. The
mass transferred into the bubble interior may easily dominate the total
mass within the bubble. \citet{Hanami:1987} used this same approach to
generalize the \citet{Weaver:1977} model to an ambient density with a
power-law slope, as well as considering the conductive evaporation of
embedded clumps. An analytical solution based on the
\citet{Weaver:1977} model for WBBs in a density gradient was obtained
by \citet{Garcia-Segura:1995}.  Similarity solutions of WBBs
mass-loaded by embedded clumps and obtained without the assumption of
a constant pressure shocked wind were described by
\citet*{Pittard:2001a} and \citet{Pittard:2001b}.
A similarity solution analogous to the
\citet{Weaver:1977} model, but taking into account the time-dependence
of the stellar wind, was obtained by \citet{Zhekov:1996}, while
\citet{Zhekov:1998} derived a similarity solution (based on
the \citet{Weaver:1977} model) with reduced thermal conduction.

\begin{figure}[t]
\centering
\includegraphics[width=0.8\textwidth]{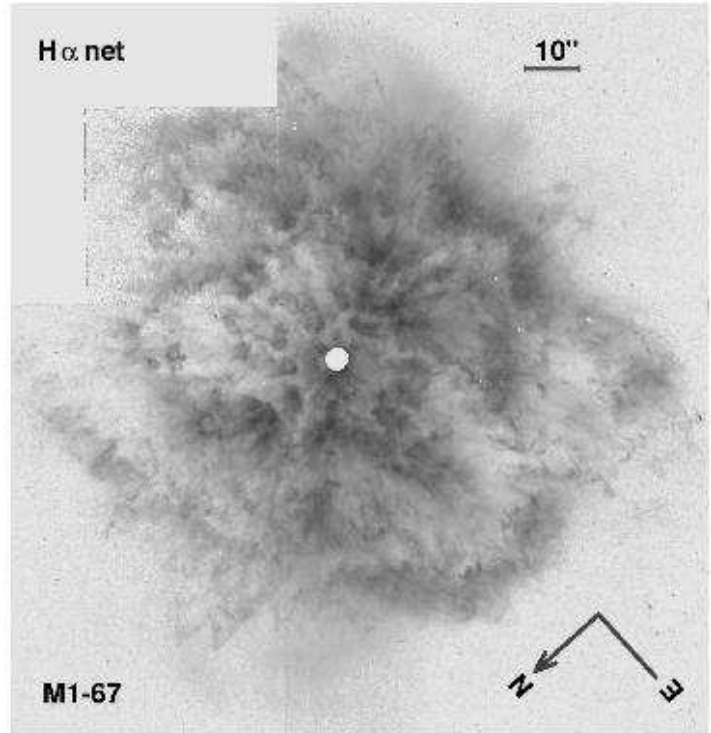}
\caption{The M1-67 nebula around WR\thinspace124. This is the youngest known WR
nebula, and unlike older nebulae there is no obvious global shell surrounding
a central cavity. Instead, the interaction between the current wind and
its LBV progenitor starts close to the central star and extends to the
outer boundary \citep[reproduced from][]{Grosdidier:1998}.}
\label{fig:m1-67}
\end{figure}

Some of the best studied WBBs are around WR stars. Many of these
display a clumpy morphology, and mass-loading is likely to be an
important process. Evidence for mass-loading in WBBs was obtained by
\citet{Smith:1984}, who observed a correlation between the velocity
and ionization potential of ultraviolet absorption features towards
the central star of the RCW~58 WBB. The slope of this relationship
effectively measures the ratio of the mass flux to the pressure in the
line-forming region, i.e. $\rho v/P$, and the value of this ratio
implies that the rate of mass loading into the bubble is 40-50 times
greater than the mass-loss rate of the WR star
\citep{Hartquist:1986}. A standard bubble without mass-loading is
unable to explain the wide range of observed velocities, while the
\citet{Weaver:1977} model predicts a correlation of the opposite
sense, irrespective of the assumed ambient density gradient \citep[see
][]{Hanami:1987}.  In contrast, similarity solutions with mass loading
from embedded clumps {\em are} able to reproduce this correlation
\citep{Hanami:1987,Pittard:2001a,Pittard:2001b}, as can hydrodynamical
simulations of the same process \citep{Arthur:1993,Arthur:1996}.
Further evidence for mass-loading in wind-blown bubbles is provided by
spectropscopic data of core-halo PNe, which indicate halo temperatures
in excess of that which can be obtained by photoionziation alone
\citep{Meaburn:1991}. \citet{Dyson:1992} and \citet{Arthur:1994} have
shown that this is consistent with a transonic wind leaving a
mass-loaded core region and shocking against clumps in the halo
region.
 
Without some form of mass-loading, the hot gas in WBBs, PNe, and
superbubbles may be too rarefied to produce observable X-ray emission.
Irrespective of the entrainment process, mass-loading increases the
interior density of the bubble while simultaneous reducing its
temperature (the same thermal energy must now be shared between more
particles), and as a consequence, the X-ray emission both softens and
dramatically increases.  We are therefore fortunate that the last
three decades have been a golden age of X-ray astronomy, with the operation
of successive facilities with vastly improved sensitivity, spatial-
and energy-resolution.  The observed temperatures in WBBs, PNe, and
quiescent (i.e. those without recent SNe) superbubbles are all
reasonably soft and therefore indicative of mass-loading (though other
explanations are also possible). However, applications of the
\citet{Weaver:1977} and \citet{Garcia-Segura:1995} models predict far
too much X-ray emission (sometimes by a factor of 100), so at the very
least the evaporation of mass from a dense swept-up shell must occur
at a much reduced rate.

Only two WBBs have been detected to date at X-ray energies, NGC~6888
\citep{Wrigge:1999} and S308 \citep{Chu:2003}, both of which are
limb-brightened. At first, this morphology seems to agree with the
predictions of the \citet{Weaver:1977} and \citet{Garcia-Segura:1995}
models since the emissivity is highest where the density is highest
and the temperature lowest.  However, the soft emission from the limb
is easily absorbed by the ISM, with the result that such models
predict a centre-filled appearance \citep{Wrigge:2005}. Again, a
reduction in the conductivity can improve the level of agreement. It
remains to be seen whether bubbles with distributed mass-loading from
embedded clumps can better reproduce the observed limb-brightening,
though this process has some support from hydrodynamical simulations
which often show instabilities breaking dense clumps off cold swept-up
shells and subsequently becoming entrained in the hot bubble interior
(e.g., see Fig.~6 in the chapter by Jane Arthur).  An alternative
explanation is that the X-ray emission from NGC~6888 and S308 arises
from the shocked RSG wind, and is currently enhanced in these objects
by a collision between the WR and RSG shells
\citep*{Freyer:2006}. However, this requires that the RSG wind is fast
($\sim 100 \;\kmps$), and speeds of this order are far from certain
\citep*[see discussion in][]{Garcia-Segura:1996}.  Finally, there is a
vast amount of evidence that the stellar winds themselves are clumpy,
which has further implications for their interaction (see, e.g.,
Fig.~\ref{fig:m1-67}).

Discerning the morphology of PNe is more difficult, as their angular
size is smaller. Only Mz~3 and NGC~6543 are adequately resolved, and,
like the WBBs around massive stars, are limb-brightened
\citep{Chu:2001,Kastner:2003}. Unlike the WBBs around massive stars,
internal absorption within the nebula is likely to be important, while
the action of collimated outflows and heat conduction may also modify
the observed X-ray morphology \citep{Kastner:2002}.  The duration for
the presence of hot gas appears to be short, as only young PNe show
diffuse X-ray emission.

On larger scales, the combined action of stellar winds and SNe from
massive stars in OB associations sweep up the ambient ISM to form
superbubbles. The physical structure of a superbubble is expected to
be similar to that of a WBB formed around an isolated massive
star. Superbubbles containing recent SNe have X-ray luminosities which
exceed the predictions from the \citet{Weaver:1977} model, but when in
a quiescent state (i.e. without recent SN blasts) the X-ray
luminosities are an order of magnitude lower than expected from the
\citet{Weaver:1977} model \citep{Chu:1995}.  In contrast to WBBs and
PNe, quiescent superbubbles have a center-filled X-ray morphology,
with brighter X-ray emission near the central star cluster, and hotter
gas ($T \sim 10^{7}\;{\rm K}$). The diffuse X-ray emission from
high-mass star-forming regions has also been studied on smaller scales
\citep[see, e.g.,][]{Townsley:2003}. In M\thinspace17, it appears that
the hot plasma is mass-loaded by a factor of $\sim 10$
\citep{Dunne:2003}. However, other processes, such as particle
acceleration, and different electron and ion temperatures may also be
important.

An interesting finding for clump-embedded, conductively-driven
mass-loading is the occurence of a negative feed-back mechanism, which
sets a maximum limit to the amount that a bubble can be mass-loaded
\citep{Pittard:2001a}.  This arises as a consequence of the dependence
of the evaporation rate on temperature and the lowering of temperature
by mass loading. Since mass-loading to primary stellar wind mass
ratios approaching those inferred for RCW\thinspace58 could not be
obtained, a key conclusion is that ablation from embedded clumps must
be the dominant driver of mass-injection in RCW\thinspace58 in
particular, and perhaps also more generally (this finding is in
agreement with the estimate that $\Mdot_{\rm ab} > \Mdot_{\rm con}$ in
Sec.~\ref{sec:mdot_comp}).  However, the lower mass-loading to primary
stellar wind mass ratios deduced for S\thinspace308 and M\thinspace17
probably do not rule out conductively-driven mass-loading in these
sources.

\subsection{SNRs and Starburst Superwinds}
The fact that the ISM is known to be multi-phase means that SNRs are
undoubtedly mass-loaded, and there is clear observational evidence of
engulfed clumps within the SNR N63A
\citep*{Chu:1999,Warren:2003}. Young SNRs first interact with the
circumstellar material ejected by their progenitor.  The youngest SNR
in the Galaxy, Cas~A, contains bright, slow-moving knots of gas called
quasi-stationary flocculi (QSFs), which have been suggested to arise
from the circumstellar bubble of the WN progenitor
\citep{Chevalier:1978}.

Similarity solutions for SNRs mass-loaded from embedded clouds have
been obtained by \citet{McKee:1977}, \citet{Chieze:1981}, and
\citet{White:1991} for conductively-driven evaporation, and by
\citet{Dyson:1987} for ablation-driven injection.  A small number of
papers based on numerical simulations of mass-loaded supernova
remnants also exist in the literature. \citet*{Cowie:1981} included the
dynamics of the clumps and found that warm clumps are swept towards
the shock front and are rapidly destroyed, while cold clumps are more
evenly distributed and have longer lifetimes.  \citet{Arthur:1996b}
studied the effects of mass loading by hydrodynamic ablation on
supernova remnants evolving inside cavities evacuated by the stellar
winds of the progenitor stars.

When SNRs overlap in regions with vigorous star formation, they may
create highly pressurized superbubbles that burst out into
intergalactic space. In the standard picture of such starburst
superwinds, the wind remains subsonic until it reaches the boundary of
the starburst region \citep{Chevalier:1985}.  However, the predicted
X-ray luminosity of the thermalized SN and stellar wind ejecta is
lower than observed unless the superwind is heavily mass-loaded
\citep{Suchkov:1996}, with the rate of mass-injection from the destruction
of clouds several times larger than that due to SNe \citep[see
also][]{Hartquist:1997}. Recently, high-spatial-resolution
observations have provided a much clearer view of the mass-loading
process in superwinds. Cool H$\alpha$ emitting filaments and clumps
are seen embedded within the superwind, and the soft X-ray emission
appears to be associated with hot gas interacting with these
structures \citep[see][and references therein]{Strickland:2004}.
Theoretical support comes from hydrodynamic models which show that 
the superwind sweeps up and incorporates large masses of material
from within the galactic disk as it develops \citep{Strickland:2000}.

Since superwinds are driven by overlapping SNRs, and the range and
radiative energy losses of a remnant are affected by mass loading, it
is desirable to have approximations which describe remnant evolution
and range in clumpy media. The first steps towards this goal were
taken by \citet*{Dyson:2002} and \citet{Pittard:2003}, where
hydrodynamical simulations of SNRs undergoing mass-loading driven
either by ablation or conduction were calculated.  Significant
differences between the evolution of the SNRs were discovered, due to
the way in which conductive mass loading is extinguished at fairly
early times, once the interior temperature of the remnant falls below
$\sim 10^{7}$~K.  At late times, remnants that ablatively mass load
are dominated by loaded mass and thermal energy, while those that
conductively mass load are dominated by swept-up mass and kinetic
energy. These works may ultimately be used in superwind models
which are akin to the \citet{McKee:1977} model of the interstellar
medium.

\subsection{AGN-SNR Interaction}
Many theoretical explanations have been proposed for the origin of the
broad emission line regions (BELR) in AGNs \citep[see, e.g.,][and
references therein]{Pittard:2003b}, including the interaction of an
AGN wind with supernovae and star clusters
\citep{Perry:1985,Williams:1994}. It is now clear that the bulk of the
broad-line emission arises from an accretion-disk wind \citep*[][see
also the chapter by Stuart Lumsden]{Proga:2000}, but interactions
between an AGN wind and SNRs may make a non-negligible contribution
to the emission of high ionization lines.  These interactions will
also mass-load the AGN wind \citep[e.g.][]{Smith:1996}, and possibly
could be used as a diagnostic of it.  Hydrodynamical simulations of
the early stages of this interaction have been presented by
\citet{Pittard:2001c} and \citet{Pittard:2003b}.  The strong radiation
field means that cool post-shock gas forms only for a very limited
time, before being heated back up to the Compton temperature.

\subsection{Intracluster Gas}
The intracluster medium consists of hot, subsonic gas which is bound
within the gravitational potential of the cluster. It was suggested
almost 3 decades ago that the density of gas within the central
regions is high enough to permit significant cooling within a Hubble
time This energy loss causes outer gas to flow subsonically in to the centre in
order to maintain hydrostatic equilibrium, a process referred to as a
`cooling flow'. Recent X-ray observations appear to show a
systematic deficit of low temperature emission in
comparison to the standard isobaric cooling-flow model, which has led
to the questioning of this model. Mass deposition rates
significantly smaller than expected are also inferred \citep[see,
e.g.,][and references therein]{Peterson:2003}.  Attention has now
focussed on the possibility that the gas is prevented from cooling by
some compensating form of energy injection.  For instance, the
frequent occurence of X-ray cavities coincident with radio lobes
around the central dominant galaxy demonstrates that AGNs can
significantly influence the X-ray morphology of the hot gas
\citep[e.g.,][]{Fabian:2002}. However, the timescale for energy
transfer from the relativistic plasma to the surrounding gas is poorly
known, while the effectiveness of heat conduction at preventing
cooling is enthusiastically debated.

\begin{figure}[t]
\centering
\includegraphics[width=8.8cm]{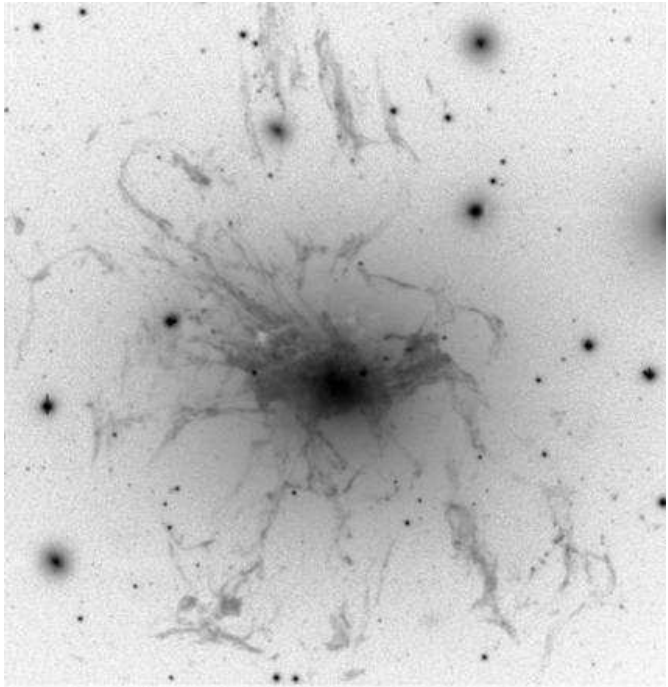}
\caption{An H$\alpha$ image showing the filamentary structure around 
NGC~1275 (credit C. Conselice (Caltech), WIYN, AURA, NOAO, NSF).}
\label{fig:ngc1275}
\end{figure}

An alternative interpretation is that the cooling plasma radiates its
energy in the UV/optical bands rather than at X-ray wavelengths.  This
can occur if there is mixing between the cooling plasma and colder
material \citep[e.g.,][]{Fabian:2002}, and may be reprocessed into the
IR if the gas is dusty. Luminous optical/UV nebulosity is common in
cluster cooling flows \citep[e.g.,][]{Heckman:1989}, and is
particularly widespread around NGC~1275 in the Perseus cluster
\citep*{Conselice:2001} (see Fig.~\ref{fig:ngc1275}).  While present
information is rather sparse, and some of this emission will be
powered by star formation, the total submillimetre to UV emission
appears to be sufficient to account for the missing soft X-ray
emission \citep{Fabian:2002}. However, direct evidence for the UV
emission resulting from this mixing is rather tentative in current
FUSE data \citep{Oegerle:2001}.

If cold clouds are embedded within the hot gas, they could be
important for global mass injection into the hot surroundings.
Indeed, since the intracluster gas is observed to be enriched with
metals, material must either be injected by the host galaxies, through
fountain flows or superwinds, or ablatively stripped from
them. Mass-loading usually reduces the mean temperature of a flow, as
the thermal energy is shared between more particles, but in clusters
this situation is modified by the gravitational forces on the flow and
the significant gravitational potential energy that the entrained
material possesses. A preliminary study has shown that under such
conditions it is possible to increase the radiated power at high
temperatures \citep{Pittard:2004}. Hence the differential luminosity
distribution can have a positive slope with $T$, as required for
agreement with the latest observations of clusters. In addition,
ripples found in the X-ray emission map of the Perseus cluster
\citep{Fabian:2003} resemble the structure of some mass-loaded
accretion flows. The most likely explanation for their origin may be
time variations in the source of outflowing material, but this
interaction clearly needs further investigation. 

\section{Summary}
The mass-loading of flows is now recognized as a fundamental process
in astronomy, yet much work remains. We still do not have a good
understanding of ablatively-driven evaporation or conductively-driven
thermal evaporation and the ability of plasma instabilities to quench
it. The rapidity with which cold material is mixed into a hotter flow
is also unknown.  The global properties of a variety of mass-loaded
flows have now been studied, but in only a few instances have specific
objects been modelled and a direct comparison with observations made.

\section*{Acknowledgements}
It is a pleasure to thank John Dyson for his encouragement and friendship,
and to my other collaborators, Tom Hartquist and Sam Falle, for theirs.


\begin{thebibliography}{}
\bibitem[\protect\citeauthoryear{Arthur, Dyson \& Hartquist}{Arthur et al.}{1993}]{Arthur:1993} 
Arthur, S.~J., Dyson, J.~E., \& Hartquist, T.~W.\ 1993, A\&A, 261, 425 
\bibitem[\protect\citeauthoryear{Arthur, Dyson \& Hartquist}{Arthur et al.}{1994}]{Arthur:1994} 
Arthur, S.~J., Dyson, J.~E., \& Hartquist, T.~W.\ 1994, A\&A, 269, 1117 
\bibitem[\protect\citeauthoryear{Arthur \& Henney}{1996}]{Arthur:1996b} 
Arthur, S.~J., \& Henney, W.~J.\ 1996, ApJ, 457, 752
\bibitem[\protect\citeauthoryear{Arthur, Henney \& Dyson}{Arthur et al.}{1996}]{Arthur:1996} 
Arthur, S.~J., Henney, W.~J., \& Dyson, J.~E.\ 1996, A\&A, 313, 897 
\bibitem[\protect\citeauthoryear{Arthur \& Lizano}{1997}]{Arthur:1997} 
Arthur, S.~J., Lizano, S.\ 1997, ApJ, 484, 810
\bibitem[\protect\citeauthoryear{Balbus}{1986}]{Balbus:1986}
Balbus, S.~A.\ 1986, ApJ, 304, 787
\bibitem[\protect\citeauthoryear{Balbus \& McKee}{1982}]{Balbus:1982}
Balbus, S.~A., \& McKee, C.~F.\ 1982, ApJ, 252, 529
\bibitem[\protect\citeauthoryear{Bally et al.}{1998}]{Bally:1998}
Bally, J., Sutherland, R.~S., Devine, D., \& Johnstone, D.\ 1998, AJ, 116, 293
\bibitem[\protect\citeauthoryear{Bandiera \& Chen}{1994a}]{Bandiera:1994a}
Bandiera, R., \& Chen, Y.\ 1994a, A\&A, 284, 629
\bibitem[\protect\citeauthoryear{Bandiera \& Chen}{1994b}]{Bandiera:1994b}
Bandiera, R., \& Chen, Y.\ 1994b, A\&A, 284, 637
\bibitem[\protect\citeauthoryear{Bertoldi}{1989}]{Bertoldi:1989}
Bertoldi, F.\ 1989, ApJ, 346, 735
\bibitem[\protect\citeauthoryear{Bertoldi \& McKee}{1990}]{Bertoldi:1990}
Bertoldi, F.\ 1990, ApJ, 354, 529
\bibitem[\protect\citeauthoryear{B\"{o}hringer \& Hartquist}{1987}]{Bohringer:1987}
B\"{o}hringer, H., \& Hartquist, T.~W.\ 1987, MNRAS, 228, 915
\bibitem[\protect\citeauthoryear{Boroson et al.}{1995}]{Boroson:1995}
Boroson, B., McCray, R., Clark, C.~O., Slavin, J., Mac~Low, M.-M., Chu, Y., \& Van Buren, D.\ 1997, ApJ, 478, 638 [Erratum: 1997, ApJ, 485, 436]
\bibitem[\protect\citeauthoryear{Cant\'{o} \& Raga}{1991}]{Canto:1991}
Cant\'{o}, J., \& Raga, A.~C.\ 1991, ApJ, 372, 646
\bibitem[\protect\citeauthoryear{Chevalier \& Clegg}{1985}]{Chevalier:1985}
Chevalier, R.~A., \& Clegg, A.~W.\ 1985, Nature, 317, 44
\bibitem[\protect\citeauthoryear{Chevalier \& Kirshner}{1978}]{Chevalier:1978}
Chevalier, R.~A., \& Kirshner, R.~P.\ 1978, ApJ, 219, 931
\bibitem[\protect\citeauthoryear{Chi\`{e}ze \& Lazareff}{1981}]{Chieze:1981}
Chi\`{e}ze, J.~P., \& Lazareff, B.\ 1981, A\&A, 95, 194
\bibitem[\protect\citeauthoryear{Cho et al.}{2003}]{Cho:2003}
Cho, J., Lazarin, A., Honein, A., Knaepen, B., Kassinos, S., \& Moin, P.\
2003, ApJ, 589, L77
\bibitem[\protect\citeauthoryear{Chu}{1982}]{Chu:1982}
Chu, Y.-H.\ 1982, ApJ, 254, 578
\bibitem[\protect\citeauthoryear{Chu et al.}{1995}]{Chu:1995}
Chu, Y.-H., Chang, H., Su, Y., \& Mac Low, M.-M.\ 1995, ApJ, 450, 157
\bibitem[\protect\citeauthoryear{Chu et al.}{2001}]{Chu:2001}
Chu, Y.-H., Guerrero, M.~A., Gruendl, R.~A., Williams, R.~M., \& Kaler, J.~B.\ 2001, ApJ, 553, L69 [Erratum: 2001, ApJ, 554, 233]
\bibitem[\protect\citeauthoryear{Chu et al.}{2003}]{Chu:2003}
Chu, Y.-H., Guerrero, M.~A., Gruendl, R.~A., Garc\'{\i}a-Segura, G., Wendker, H.~J.\ 2003, ApJ, 599, 1189
\bibitem[\protect\citeauthoryear{Chu et al.}{1999}]{Chu:1999}
Chu, Y.-H., et al.\ 1999, New Views of the Magellanic Clouds, eds.
Y.-H.~Chu, N.~Suntzeff, J.~Hesser, \& D.~Bohlender, IAU~Symp., 190, 143
\bibitem[\protect\citeauthoryear{Conselice, Gallagher \& Wyse}{Conselice et al.}{2001}]{Conselice:2001} 
Conselice, C.~J., Gallagher, J.~S., \& Wyse, R.~F.~G.\ 2001, AJ, 122, 2281
\bibitem[\protect\citeauthoryear{Cowie \& McKee}{1977}]{Cowie:1977}
Cowie, L.~L., \& McKee, C.~F.\ 1977, ApJ, 211, 135
\bibitem[\protect\citeauthoryear{Cowie, McKee \& Ostriker}{Cowie et al.}{1981}]{Cowie:1981}
Cowie, L.~L., McKee, C.~F., \& Ostriker, J.~P.\ 1981, ApJ, 247, 908
\bibitem[\protect\citeauthoryear{Cowie \& Songalia}{1977}]{Cowie:1977b}
Cowie, L.~L., \& Songalia, A.\ 1977, Nature, 266, 501
\bibitem[\protect\citeauthoryear{David \& Durisen}{1989}]{David:1989}
David, L.~P., \& Durisen, R.~H.\ 1989, ApJ, 346, 618
\bibitem[\protect\citeauthoryear{Draine \& Giuliani}{1984}]{Draine:1984}
Draine, B.~T., \& Giuliani, J.~L.\ 1984, ApJ, 281, 690
\bibitem[\protect\citeauthoryear{Dunne et al.}{2003}]{Dunne:2003}
Dunne, B.~C., et al.\ 2003, ApJ, 590, 306
\bibitem[\protect\citeauthoryear{Durisen \& Burns}{1981}]{Durisen:1981}
Durisen, R.~H., \& Burns, J.~O.\ 1981, MNRAS, 195, 535
\bibitem[\protect\citeauthoryear{Dyson}{1968}]{Dyson:1968}
Dyson, J.~E.\ 1968, Ap\&SS, 1, 388
\bibitem[\protect\citeauthoryear{Dyson}{1973}]{Dyson:1973}
Dyson, J.~E.\ 1973, A\&A, 23, 381
\bibitem[\protect\citeauthoryear{Dyson}{1992}]{Dyson:1992}
Dyson, J.~E.\ 1992, MNRAS, 255, 460
\bibitem[\protect\citeauthoryear{Dyson}{1994}]{Dyson:1994}
Dyson, J.~E.\ 1994, in Lecture Notes in Physics, 431, Star Formation Techniques
in Infrared and mm-Wave Astronomy, ed. T.~P.~Ray \& S.~V.~W. Beckwith 
(Berlin:Springer), 93
\bibitem[\protect\citeauthoryear{Dyson, Arthur \& Hartquist}{Dyson et al.}{2002}]{Dyson:2002}
Dyson, J.~E., Arthur, S.~J., \& Hartquist, T.~W.\ 2002, A\&A, 390, 1063
\bibitem[\protect\citeauthoryear{Dyson \& de Vries}{1972}]{Dyson:1972}
Dyson, J.~E., \& de Vries, J.\ 1972, A\&A, 20, 223
\bibitem[\protect\citeauthoryear{Dyson \& Hartquist}{1987}]{Dyson:1987}
Dyson, J.~E., \& Hartquist, T.~W.\ 1987, MNRAS, 228, 453
\bibitem[\protect\citeauthoryear{Dyson \& Hartquist}{1994}]{Dyson:1994b}
Dyson, J.~E., \& Hartquist, T.~W.\ 1994, MNRAS, 269, 447
\bibitem[\protect\citeauthoryear{Dyson et al.}{1989}]{Dyson:1989}
Dyson, J.~E., Hartquist, T.~W., Pettini, M., \& Smith, L.~J.\ 1989, MNRAS, 241, 625
\bibitem[\protect\citeauthoryear{Dyson, Williams \& Redman}{Dyson et al.}{1995}]{Dyson:1995}
Dyson, J.~E., Williams, R.~J.~R., \& Redman, M.~P.\ 1995, MNRAS, 237, 700
\bibitem[\protect\citeauthoryear{Elmegreen \& Lada}{1977}]{Elmegreen:1977}
Elmegreen, B.~G., \& Lada, C.~J.\ 1977, ApJ, 214, 725
\bibitem[\protect\citeauthoryear{Fabian et al.}{2002}]{Fabian:2002}
Fabian, A.~C., Allen, S.~W., Crawford, C.~S., Johnstone, R.~W., Morris, R.~G., 
Sanders, J.~S., \& Schmidt, R.~W.\ 2002, MNRAS, 332, L50
\bibitem[\protect\citeauthoryear{Fabian et al.}{2003}]{Fabian:2003}
Fabian, A.~C.,et al.\ 2003, MNRAS, 344, L43
\bibitem[\protect\citeauthoryear{Falle}{1975}]{Falle:1975}
Falle, S.~A.~E.~G\ 1975, A\&A, 43, 323
\bibitem[\protect\citeauthoryear{Falle et al.}{2002}]{Falle:2002}
Falle, S.~A.~E.~G, Coker, R.~F., Pittard, J.~M., Dyson, J.~E., \& Hartquist, T.~W.\ 2002, MNRAS, 329, 670
\bibitem[\protect\citeauthoryear{Ferrara \& Shchekinov}{1993}]{Ferrara:1993}
Ferrara, A., \& Shchekinov, Y.\ 1993, ApJ, 417, 595
\bibitem[\protect\citeauthoryear{Field}{1965}]{Field:1965}
Field, G.~B.\ 1965, ApJ, 142, 531
\bibitem[\protect\citeauthoryear{Fragile et al.}{2005}]{Fragile:2005}
Fragile, P.~C., Anninos, P., Gustafson, K., \& Murray, S.~D.\ 2005, ApJ, 619, 327
\bibitem[\protect\citeauthoryear{Fragile et al.}{2004}]{Fragile:2004}
Fragile, P.~C., Murray, S.~D., Anninos, P., \& Breugel, W.~V.\ 2004, ApJ, 604, 74
\bibitem[\protect\citeauthoryear{Freyer, Hensler \& Yorke}{Freyer et al.}{2006}]{Freyer:2006}
Freyer, T., Hensler, G., \& Yorke, H.~W.\ 2006, ApJ, 638, 262
\bibitem[\protect\citeauthoryear{Galeev \& Natanzon}{1984}]{Galeev:1984}
Galeev, A.~A., \& Natanzon, A.~M.\ 1984, Dokl. Phys., 275, 6
\bibitem[\protect\citeauthoryear{Garc\'{\i}a-Segura, Langer \& Mac Low}{Garc\'{\i}a-Segura et al.}{1996}]{Garcia-Segura:1996}
Garc\'{\i}ia-Segura, G., Langer, N., \& Mac Low, M.~M.\ 1996, A\&A, 316, 133
\bibitem[\protect\citeauthoryear{Garc\'{\i}a-Segura \& Mac Low}{1995}]{Garcia-Segura:1995}
Garc\'{\i}ia-Segura, G., \& Mac Low, M.~M.\ 1995, ApJ, 455, 145
\bibitem[\protect\citeauthoryear{Gary \& Feldman}{1977}]{Gary:1977}
Gary, S.~P., \& Feldman, W.~C.\ 1977, J. Geophs. Res., 82, 1087
\bibitem[\protect\citeauthoryear{Giuliani}{1984}]{Giuliani:1984}
Giuliani, J.~L.\ 1984, ApJ, 277, 605
\bibitem[\protect\citeauthoryear{Grosdidier et al.}{1998}]{Grosdidier:1998}
Grosdidier, Y., Moffat, A.~F.~J., Joncas, G., Acker, A.\ 1998, ApJ, 506, L127
\bibitem[\protect\citeauthoryear{Hanami \& Sakashita}{1987}]{Hanami:1987}
Hanami, H., \& Sakashita, S.\ 1987, A\&A, 181, 343
\bibitem[\protect\citeauthoryear{Hartquist \& Dyson}{1988}]{Hartquist:1988} 
Hartquist, T.~W., \& Dyson, J.~E.\ 1988, Ap\&SS, 144, 615
\bibitem[\protect\citeauthoryear{Hartquist et al.}{1986}]{Hartquist:1986} 
Hartquist, T.~W., Dyson, J.~E., Pettini, M., \& Smith, L.~J.\ 1986, 
MNRAS, 221, 715
\bibitem[\protect\citeauthoryear{Hartquist \& Dyson}{1993}]{Hartquist:1993} 
Hartquist, T.~W., \& Dyson, J.~E.\ 1993, QJRAS, 34, 57
\bibitem[\protect\citeauthoryear{Hartquist, Dyson \& Williams}{Hartquist et al.}{1997}]{Hartquist:1997} 
Hartquist, T.~W., Dyson, J.~E., \& Williams, R.~J.~R.\ 1997, ApJ, 482, 182
\bibitem[\protect\citeauthoryear{Heckman et al.}{1989}]{Heckman:1989}
Heckman, T.~M., Baum, S.~A., van Breugel, W.~J.~M., \& McCarthy, P.\ 1989, 
ApJ, 338, 48
\bibitem[\protect\citeauthoryear{Hensler \& Vieser}{2002}]{Hensler:2002}
Hensler, G., \& Vieser, W.\ 2002, Ap\&SS, 281, 275
\bibitem[\protect\citeauthoryear{Jun, Jones \& Norman}{Jun et al.}{1996}]{Jun:1996}
Jun, B.-I., Jones, T.~W., \& Norman, M.~L.\ 1996, ApJ, 468, L59
\bibitem[\protect\citeauthoryear{Kahn}{1969}]{Kahn:1969}
Kahn, F.~D.\ 1969, Physica, 41, 172
\bibitem[\protect\citeauthoryear{Kastner et al.}{2003}]{Kastner:2003}
Kastner, J.~H., Balick, B., Blackman, E.~G., Frank, A., Soker, N., Vrt\'{\i}lek, S.~D., \& Li, J.\ 2003, ApJ, 591, L37
\bibitem[\protect\citeauthoryear{Kastner et al.}{2002}]{Kastner:2002}
Kastner, J.~H., Li, J., Vrt\'{\i}lek, S.~ D., Gatley, I., Merrill, K.~M., \& Soker, N.\ 2002, 581, 1225
\bibitem[\protect\citeauthoryear{Klein et al.}{2003}]{Klein:2003}
Klein, R.~I., Budil, K.~S., Perry, T.~S., Bach, D.~R.\ 2003, ApJ, 583, 245
\bibitem[\protect\citeauthoryear{Klein, McKee \& Colella}{Klein et al.}{1994}]{Klein:1994}
Klein, R.~I., McKee, C.~F., \& Colella, P.\ 1994, ApJ, 420, 213
\bibitem[\protect\citeauthoryear{Landau \& Lifshitz}{1959}]{Landau:1959} 
Landau, L.~D., \& Lifshitz, E.~M.\ 1959. Fluid Mechanics, Pergamon Press, 
Oxford
\bibitem[\protect\citeauthoryear{Lefloch \& Lazareff}{1994}]{Lefloch:1994}
Lefloch, B., \& Lazareff, B.\ 1994, A\&A, 289, 559
\bibitem[\protect\citeauthoryear{Levinson \& Eichler}{1992}]{Levinson:1992}
Levinson, A., \& Eichler, D.\ 1992, ApJ, 387, 212
\bibitem[\protect\citeauthoryear{McKee \& Cowie}{1975}]{McKee:1975}
McKee, C.~F., \& Cowie, L.~L.\ 1975, ApJ, 195, 715 
\bibitem[\protect\citeauthoryear{McKee \& Cowie}{1977}]{McKee:1977b}
McKee, C.~F., \& Cowie, L.~L.\ 1977, ApJ, 215, 213
\bibitem[\protect\citeauthoryear{McKee \& Ostriker}{1977}]{McKee:1977} 
McKee, C.~F., \& Ostriker, J.~P.\ 1977, ApJ, 218, 148
\bibitem[\protect\citeauthoryear{Marcolini et al.}{2005}]{Marcolini:2005}
Marcolini, A., Strickland, D.~K., D'Ercole, A., Heckman, T.~M., Hoopes, C.~G.\ 2005, MNRAS, 362, 626 
\bibitem[\protect\citeauthoryear{Meaburn et al.}{2005}]{Meaburn:2005}
Meaburn, J., Boumis, P., L\'{o}pez, J.~A., Harman, D.~J., Bryce, M., Redman, M.~P., \& Mavromatakis, F.\ 2005, MNRAS, 360, 963
\bibitem[\protect\citeauthoryear{Meaburn et al.}{1996}]{Meaburn:1996}
Meaburn, J., Clayton, C.~A., Bryce, M., \& Walsh, J.~R.\ 1996, MNRAS, 281, L57
\bibitem[\protect\citeauthoryear{Meaburn et al.}{1991}]{Meaburn:1991}
Meaburn, J., Nicholson, R.~A., Bryce, M., Dyson, J.~E., \& Walsh, J.~R.\ 1991, MNRAS, 252, 535
\bibitem[\protect\citeauthoryear{Mellema, Kurk \& R\"{o}ttgering}{Mellema et al.}{2002}]{Mellema:2002}
Mellema, G., Kurk, J.~D., \& R\"{o}ttgering, H.~J.~A.\ 2002, A\&A, 395, L13
\bibitem[\protect\citeauthoryear{Mellema et al.}{1998}]{Mellema:1998}
Mellema, G., Raga, A.~C., Cant\'{o}, J., Lundqvist, P., Balick, B., 
Steffen, W., \& Noriega-Crespo, A.\ 1998, A\&A, 331, 335
\bibitem[\protect\citeauthoryear{O'Dell, Wen \& Hu}{O'Dell et al.}{1993}]{ODell:1993}
O'Dell, C.~R., Wen, Z., \& Hu, X.\ 1993, ApJ, 410, 696
\bibitem[\protect\citeauthoryear{Oegerle et al.}{2001}]{Oegerle:2001} 
Oegerle, W.~R., et al.\ 2001, ApJ, 560, 187
\bibitem[\protect\citeauthoryear{Orlando et al.}{2005}]{Orlando:2005}
Orlando, S., Peres, G., Reale, F., Bocchino, F., Rosner, R., Plewa, T., Siegel, A.\ 2005, A\&A, 444, 505
\bibitem[\protect\citeauthoryear{Pavlakis et al.}{2001}]{Pavlakis:2001}
Pavlakis, K.~G., Williams, R.~J.~R., Dyson, J.~E., Falle, S.~A.~E.~G.,
\& Hartquist, T.~W.\ 2001, A\&A, 369, 263
\bibitem[\protect\citeauthoryear{Perry \& Dyson}{1985}]{Perry:1985}
Perry, J.~J., \& Dyson, J.~E.\ 1985, MNRAS, 213, 665
\bibitem[\protect\citeauthoryear{Peterson et al.}{2003}]{Peterson:2003}
Peterson, J.~R., et al.\ 2003, ApJ, 590, 207
\bibitem[\protect\citeauthoryear{Pittard et al.}{2003}]{Pittard:2003}
Pittard, J.~M., Arthur, S.~J., Dyson, J.~E., Falle, S.~A.~E.~G., Hartquist,
T.~W., Knight M.~I., \& Pexton M.\ 2003, A\&A, 401, 1027
\bibitem[\protect\citeauthoryear{Pittard, Dyson \& Hartquist}{Pittard et al.}{2001a}]{Pittard:2001a}
Pittard, J.~M., Dyson, J.~E., \& Hartquist, T.~W.\ 2001a, A\&A, 367, 1000
\bibitem[\protect\citeauthoryear{Pittard et al.}{2001c}]{Pittard:2001c}
Pittard, J.~M., Dyson, J.~E., Falle, S.~A.~E.~G., \& Hartquist, T.~W.\ 2001c, A\&A, 375, 827
\bibitem[\protect\citeauthoryear{Pittard et al.}{2003b}]{Pittard:2003b}
Pittard, J.~M., Dyson, J.~E., Falle, S.~A.~E.~G., \& Hartquist, T.~W.\ 2003b, A\&A, 408, 79
\bibitem[\protect\citeauthoryear{Pittard et al.}{2005}]{Pittard:2005}
Pittard, J.~M., Dyson, J.~E., Falle, S.~A.~E.~G., \& Hartquist, T.~W.\ 2005, MNRAS, 361, 1077
\bibitem[\protect\citeauthoryear{Pittard, Hartquist \& Dyson}{Pittard et al.}{2001b}]{Pittard:2001b}
Pittard, J.~M., Hartquist, T.~W., \& Dyson, J.~E.\ 2001b, A\&A, 373, 1043
\bibitem[\protect\citeauthoryear{Pittard et al.}{2004}]{Pittard:2004}
Pittard, J.~M., Hartquist, T.~W., Ashmore, I., Byfield, A., Dyson, J.~E.,
\& Falle, S.~A.~E.~G.\ 2004, A\&A, 414, 399
\bibitem[\protect\citeauthoryear{Poludnenko, Frank \& Blackman}{Poludnenko et al.}{2002}]{Poludnenko:2002}
Poludnenko, A.~Y., Frank, A., \& Blackman, E.~G.\ 2002, ApJ, 576, 832
\bibitem[\protect\citeauthoryear{Proga, Stone \& Kallman}{Proga et al.}{2000}]{Proga:2000}
Proga, D., Stone, J.~M., \& Kallman, T.~R.\ 2000, ApJ, 543, 686
\bibitem[\protect\citeauthoryear{Rosner \& Tucker}{1989}]{Rosner:1989}
Rosner, R., \& Tucker, W.~H.\ 1989, ApJ, 338, 761
\bibitem[\protect\citeauthoryear{Sandford, Whitaker \& Klein}{Sandford et al.}{1982}]{Sandford:1982}
Sandford, M.~T. II, Whitaker, R.~W., \& Klein, R.~I.\ 1982, ApJ, 260, 183
\bibitem[\protect\citeauthoryear{Serabyn, Lacy \& Achtermann}{Serabyn et al.}{1991}]{Serabyn:1991}
Serabyn, E., Lacy, J.~H., \& Achtermann, J.~M.\ 1991, ApJ, 378, 557
\bibitem[\protect\citeauthoryear{Smith et al.}{1984}]{Smith:1984} 
Smith, L.~J., Pettini, M., Dyson, J.~E., \& Hartquist, T.~W.\ 1984, 
MNRAS, 211, 679
\bibitem[\protect\citeauthoryear{Smith et al.}{1988}]{Smith:1988} 
Smith, L.~J., Pettini, M., Dyson, J.~E., \& Hartquist, T.~W.\ 1988, 
MNRAS, 234, 625
\bibitem[\protect\citeauthoryear{Smith}{1996}]{Smith:1996} 
Smith, S.~J.\ 1996, ApJ, 473, 773
\bibitem[\protect\citeauthoryear{Spitzer}{1962}]{Spitzer:1962}
Spitzer, L.\ 1962. Physics of Fully Ionized Gases, Interscience, New York
\bibitem[\protect\citeauthoryear{Spitzer \& H\"{a}rm}{1953}]{Spitzer:1953}
Spitzer, L., \& H\"{a}rm, R.\ 1953, Phys. Rev., 89, 977
\bibitem[\protect\citeauthoryear{Steffen \& L\'{o}pez}{2004}]{Steffen:2004}
Steffen, W., \& L\'{o}pez, J.~A.\ 2004, ApJ, 612, 319
\bibitem[\protect\citeauthoryear{Stone \& Norman}{1992}]{Stone:1992}
Stone, J.~M., \& Norman, M.~L.\ 1992, ApJ, 390, L17
\bibitem[\protect\citeauthoryear{Strickland et al.}{2004}]{Strickland:2004}
Strickland, D.~K., Heckman, T.~M., Colbert, E.~J.~M., Hoopes, C.~G., 
\& Weaver, K.~A.\ 2004, ApJSS, 151, 193
\bibitem[\protect\citeauthoryear{Strickland \& Stevens}{2000}]{Strickland:2000}
Strickland, D.~K., \& Stevens, I.~R.\ 2000, MNRAS, 314, 511
\bibitem[\protect\citeauthoryear{Suchkov et al.}{1996}]{Suchkov:1996}
Suchkov, A.~A., Berman, V.~G., Heckman, T.~M., Balsara, D.~S.\ 1996, ApJ, 
463, 528
\bibitem[\protect\citeauthoryear{Toniazzo, Hartquist \& Durisen}{Toniazzo et al.}{2001}]{Toniazzo:2001}
Toniazzo, T., Hartquist, T.~W., \& Durisen, R.~H.\ 2001, MNRAS, 322, 149
\bibitem[\protect\citeauthoryear{Townsley et al.}{2003}]{Townsley:2003}
Townsley, L.~K., Feigelson, E.~D., Montmerle, T., Broos, P.~S., Chu, Y.-H., 
\& Garmire, G.~P.\ 2003, ApJ, 593, 874
\bibitem[\protect\citeauthoryear{Vieser \& Hensler}{2000}]{Vieser:2000}
Vieser, W., \& Hensler, G.\ 2000, Ap\&SS, 272, 189
\bibitem[\protect\citeauthoryear{Warren, Hughes \& Slane}{Warren et al.}{2003}]{Warren:2003}
Warren, J.~S., Hughes, J.~P., \& Slane, P.O.\ 2003, ApJ, 583, 260
\bibitem[\protect\citeauthoryear{Weaver et al.}{1977}]{Weaver:1977} 
Weaver, R., McCray, R., Castor, J., Shapiro, P., \& Moore, R.\ 1977, ApJ, 
218, 377
\bibitem[\protect\citeauthoryear{White \& Long}{1991}]{White:1991}
White, R.~L., \& Long, K.~S.\ 1991, ApJ, 373, 543
\bibitem[\protect\citeauthoryear{Williams}{2002}]{Williams:2002}
Williams, R.~J.~R.\ 2002, MNRAS, 331, 693
\bibitem[\protect\citeauthoryear{Williams \& Dyson}{2002}]{Williams:2002b}
Williams, R.~J.~R., \& Dyson, J.~E.\ 2002, MNRAS, 333, 1
\bibitem[\protect\citeauthoryear{Williams, Dyson \& Hartquist}{Williams et al.}{1999}]{Williams:1999}
Williams, R.~J.~R., Dyson, J.~E., \& Hartquist, T.~W.\ 1999, A\&A, 344, 675
\bibitem[\protect\citeauthoryear{Williams, Dyson \& Hartquist}{Williams et al.}{2000}]{Williams:2000}
Williams, R.~J.~R., Dyson, J.~E., \& Hartquist, T.~W.\ 2000, MNRAS, 314, 315
\bibitem[\protect\citeauthoryear{Williams, Hartquist \& Dyson}{Williams et al.}{1995}]{Williams:1995}
Williams, R.~J.~R., Hartquist, T.~W., \& Dyson, J.~E.\ 1995, ApJ, 446, 759
\bibitem[\protect\citeauthoryear{Williams \& Perry}{1994}]{Williams:1994}
Williams, R.~J.~R., \& Perry, J.~J.\ 1994, MNRAS, 269, 538
\bibitem[\protect\citeauthoryear{Wrigge}{1999}]{Wrigge:1999}
Wrigge, M.\ 1999, A\&A, 343, 599
\bibitem[\protect\citeauthoryear{Wrigge et al.}{2005}]{Wrigge:2005}
Wrigge, M., Chu, Y.-H., Magnier, E.~A., \& Wendker, H.~J.\ 2005, ApJ, 633, 248
\bibitem[\protect\citeauthoryear{Xu \& Stone}{1995}]{Xu:1995}
Xu, J., \& Stone, J.~M.\ 1995, ApJ, 454, 172
\bibitem[\protect\citeauthoryear{Yusef-Zadeh \& Melia}{1992}]{Yusef-Zadeh:1992}
Yusef-Zadeh, F., \& Melia, F.\ 1992, ApJ, 385, L41
\bibitem[\protect\citeauthoryear{Yusef-Zadeh \& Morris}{1991}]{Yusef-Zadeh:1991}
Yusef-Zadeh, F., \& Morris, M.\ 1991, ApJ, 371, L59
\bibitem[\protect\citeauthoryear{Zhekov \& Myasnikov}{1998}]{Zhekov:1998}
Zhekov, S.~A., \& Myasnikov, A.~V.\ 1998, New Ast., 3, 57
\bibitem[\protect\citeauthoryear{Zhekov \& Perinotto}{1996}]{Zhekov:1996}
Zhekov, S.~A., \& Perinotto, M.\ 1996, A\&A, 309, 648
\end{thebibliography}
\end{document}